\documentclass{JHEP3}
\usepackage{epsfig,amsmath,amssymb}

\def\vvec#1{\mbox{\bf#1}}
\def\mat#1{\mbox{\bf#1}}
\def\av#1{\langle #1 \rangle}
\def\LSfun{\Sigma}
\def\scale{{\rm scale}}
\def\stat{{\rm stat}}

\def\itB{\it}
\def\YY{Y}
\def\half{\frac{1}{2}}

\def\lN{l_{\rm n}}
\def\lF{l_{\rm f}}
\def\qN{q_{\rm n}}
\def\qF{q_{\rm f}}
\def\bN{b_{\rm n}}
\def\bF{b_{\rm f}}
\def\NO{\tilde\chi^0_1} 
\def\NT{\tilde\chi^0_2}

\def\lR{\tilde l_R} 
\def\qL{\tilde q_L}
\def\gl{\tilde g}
\def\sq{\tilde q}

\def\sb{\tilde b}

\def\bT{\tilde b_2}

\def\bO{\tilde b_1}
\def\tO{\tilde t_1}
\def\uL{\tilde u_L}

\def\mNO{m_{\NO}}
\def\mlR{m_{\lR}}
\def\mNT{m_{\NT}}
\def\mqL{m_{\qL}}
\def\mgl{m_{\gl}}

\def\mbO{m_{\bO}}

\def\msq{m_{\sq}}

\def\sNO{\mNO^2}
\def\slR{\mlR^2}
\def\sNT{\mNT^2}
\def\sqL{\mqL^2}
\def\sgl{\mgl^2}

\def\low{{\rm low}}
\def\high{{\rm high}}
\def\equal{{\rm eq}}

\def\max{{\rm max}}
\def\min{{\rm min}}
\def\GeV{{\rm GeV}}
\def\mHalf{m_{1/2}}
\def\mZero{m_0}
\def\AZero{A_0}
\def\rglqL{\frac{\mgl}{\mqL}}
\def\rqLNT{\frac{\mqL}{\mNT}}
\def\rNTlR{\frac{\mNT}{\mlR}}
\def\rlRNO{\frac{\mlR}{\mNO}}
\def\theory{{\rm th}}
\def\exp{{\rm exp}}
\def\mll{m_{ll}}
\def\mqll{m_{qll}}
\def\mqlN{m_{q\lN}}
\def\mqlF{m_{q\lF}}
\def\mql{m_{ql}}

\def\mqlHigh{m_{ql(\high)}}

\def\mqFllThres{m_{\qF ll(\theta>\frac{\pi}{2})}}
\def\mbFllThres{m_{\bF ll(\theta>\frac{\pi}{2})}}
\def\mqqll{m_{qqll}}
\def\mqq{m_{qq}}
\def\mqNl{m_{\qN l}}
\def\mqNlN{m_{\qN\lN}}
\def\mqNlF{m_{\qN\lF}}
\def\mqql{m_{qql}}
\def\mqqlN{m_{qq\lN}}
\def\mqqlF{m_{qq\lF}}
\def\mqNll{m_{\qN ll}}
\def\mqFll{m_{\qF ll}}
\def\mqFl{m_{\qF l}}
\def\mqFlN{m_{\qF\lN}}
\def\mqFlF{m_{\qF\lF}}
\def\mqFlLow{m_{\qF l(\low)}}
\def\mqFlHigh{m_{\qF l(\high)}}
\def\mqNlLow{m_{\qN l(\low)}}
\def\mqNlHigh{m_{\qN l(\high)}}
\def\mqqlLow{m_{qql(\low)}}
\def\mqqlHigh{m_{qql(\high)}}
\def\mqqlEq{m_{qql(\equal)}}
\def\mqllLow{m_{qll(\low)}}
\def\mqFlEq{m_{\qF l(\equal)}}
\def\maxmqFlEq{\mqFlEq^\max}
\def\maxmll{\mll^\max}

\def\minmqFllThres{\mqFllThres^\min}
\def\minmbFllThres{\mbFllThres^\min}
\def\maxmqqll{\mqqll^\max}
\def\maxmqq{\mqq^\max}
\def\maxmqFlN{\mqFlN^\max}
\def\maxmqFlF{\mqFlF^\max}
\def\maxmqNlN{\mqNlN^\max}
\def\maxmqNlF{\mqNlF^\max}
\def\maxmqqlN{\mqqlN^\max}
\def\maxmqqlF{\mqqlF^\max}
\def\maxmqNll{\mqNll^\max}
\def\maxmqFll{\mqFll^\max}
\def\maxmqFlLow{\mqFlLow^\max}
\def\maxmqFlHigh{\mqFlHigh^\max}
\def\maxmqqlLow{\mqqlLow^\max}
\def\maxmqqlEq{\mqqlEq^\max}
\def\maxmqllLow{\mqllLow^\max}
\def\mqqlEqCrit{\mqqlEq^{\rm crit}}
\def\mqqlEqCosB{\mqqlEq^{\cos\alpha=\pm1}}
\def\mqqlEqP{\mqqlEq^{\uparrow\uparrow}}
\def\mqqlEqAP{\mqqlEq^{\uparrow\downarrow}}
\def\pqq{|\vvec{p}_{qq}|}
\def\pqqS{\vvec{p}_{qq}}

\def\pqqP{|\pqqS^{\uparrow\uparrow}|}
\def\pqqAP{|\pqqS^{\uparrow\downarrow}|}
\def\pqqCrit{|\pqqS^{\rm crit}|}
\def\pqqCosB{|\pqqS^{\cos\alpha=\pm1}|}

\def\mbNl{m_{\bN l}}

\def\mbb{m_{bb}}
\def\mbbll{m_{bbll}}
\def\mbblLow{m_{bbl(\low)}}
\def\mbblHigh{m_{bbl(\high)}}
\def\mbNll{m_{\bN ll}}
\def\mbNlLow{m_{\bN l(\low)}}
\def\mbNlHigh{m_{\bN l(\high)}}

\def\maxmbb{\mbb^\max}
\def\maxmbbll{\mbbll^\max}
\def\maxmbblLow{\mbblLow^\max}
\def\maxmbblHigh{\mbblHigh^\max}
\def\maxmbNll{\mbNll^\max}
\def\maxmbNlLow{\mbNlLow^\max}
\def\maxmbNlHigh{\mbNlHigh^\max}

\def\SPSOa{SPS~1a}

\def\REFqFll{Eq.~(4.4) of \cite{Gjelsten:2004}}
\def\REFqlLowHigh{Eq.~(4.5) of \cite{Gjelsten:2004}}
\def\REFqlEq{Eq.~(4.8) of \cite{Gjelsten:2004}}

\allowdisplaybreaks

\title{Measurement of the Gluino Mass via Cascade Decays for SPS~1a}
\author{B.~K.~Gjelsten\\ 
Department of Physics, 
University of Oslo, 
N-0316 Oslo, Norway\\ 
E-mail: \email{B.K.Gjelsten@fys.uio.no}}
\author{D.~J.~Miller\\ 
Department of Physics and Astronomy, 
University of Glasgow, 
Glasgow G12 8QQ, 
U.K. 
E-mail: \email{D.Miller@physics.gla.ac.uk}}
\author{P.~Osland\\ 
Department of Physics, 
University of Bergen, N-5007 Bergen, Norway\\ 
and TH Division, Physics Department, CERN, CH~1211 Gen\`eve, Switzerland \\
E-mail: \email{Per.Osland@ift.uib.no}}
\abstract{If R-parity conserving supersymmetry is realised with 
masses below the TeV scale, sparticles will be produced and decay 
in cascades at the LHC. In the case of a neutral LSP, which will 
not be detected, decay chains cannot be fully reconstructed, complicating  
the mass determination of the new particles. 
In this paper we extend the method of obtaining masses from kinematical 
endpoints to include a gluino at the head of a five-sparticle decay chain. 
This represents a non-trivial extension of the corresponding method
for the squark decay chain.
We calculate the endpoints of the new distributions and assess their 
applicability by examining the theoretical distributions for a variety 
of mass scenarios. 
The precision with which the gluino mass can be determined by this method 
is investigated for the mSUGRA point SPS~1a. 
Finally we estimate the improvement obtained from adding a Linear Collider 
measurement of the LSP mass. 
}
\preprint{ATL-PHYS-2005-001\\
CERN-PH-TH/2005-098}

\begin{document}

\section{Introduction}\label{sect:intro}

Supersymmetry~\cite{Fayet:1976cr,Dimopoulos:1981zb,Nilles:1983ge,Haber:1984rc} 
provides one of the more popular extensions to the Standard Model 
at higher energies. 
It has many attractive features, one of which is a possible solution 
to the hierarchy problem~\cite{Weinberg:1975gm}. 
For this to be the case supersymmetric particles must exist 
near the TeV-scale, and may therefore be accessed by the LHC. 

Conservation of R-parity means that any interaction vertex 
must involve an even number of supersymmetric fields. 
For collider experiments this has two important consequences:
sparticles will be produced in pairs, and 
the decay chain of a sparticle will always end with the lightest
supersymmetric particle (LSP). 
A SUSY event at the LHC will then result in two decay chains, 
each giving an undetected LSP together with a number of 
detectable Standard Model particles. 
The escaping LSPs make it difficult to fully reconstruct events. 
This means that the masses of the sparticles involved in a 
decay chain cannot be readily reconstructed from the end products. 
However, even though they do not represent any particular sparticle, 
the various mass distributions that can be constructed from the 
detectable particles of the decay do have a sensitivity to the 
sparticle masses.
In particular, the maximum value of each of these distributions 
can in principle be calculated for any given decay chain, and this 
maximum value will be determined by the masses involved in the decay. 
If these kinematical endpoints are measured, it is then possible 
to obtain the sparticle masses in a numerical fit. 
This `endpoint method' has been widely used to determine masses of 
supersymmetric particles \cite{Hinchliffe:1996iu,Hinchliffe:1998zj,
Bachacou:1999zb,Polesello,Allanach:2000kt,Lester}.
The decay chain
\begin{equation}
\label{eq:squarkchain}
\qL\to\NT\qF\to\lR\lN\qF\to\NO\lF\lN\qF
\end{equation}
is particularly amenable to this method\footnote{Maintaining the
standard convention, the subscripts `n' for {\it near} and `f' for
{\it far} on the leptons and quarks are used to distinguish the order
of particle emission in the decay chain. These subscripts will be
suppressed if there is no ambiguity. In \cite{Gjelsten:2004}, which will 
be frequently referred to throughout the text, no subscript is 
used for the quark as there is only one. In the current notation this quark 
corresponds to $\qF$.}.
In \cite{Gjelsten:2004} we investigated this decay chain in considerable
detail for two points on the SPS~1a line \cite{Allanach:2002nj}: 
$(\alpha)$, which is 
the standard point and often simply referred to as `the SPS~1a point', 
and $(\beta)$, which lies at somewhat higher masses. 
In this paper we restrict ourselves to the standard SPS~1a point 
and investigate the situation where a gluino 
is at the head of the decay chain, 
\begin{equation}
\label{eq:gluinochain}
\gl\to\qL\qN\to\NT\qF\qN\to\lR\lN\qF\qN\to\NO\lF\lN\qF\qN 
\end{equation}
Seven more distributions become available with the inclusion of $\qN$,  
and their endpoints, if measurable, will enable us to find the 
gluino mass. 
For ease of reference the new distributions/endpoints will be 
called `gluino distributions/endpoints' as opposed to the 
`squark distributions/endpoints' which do not involve the gluino mass.  

Although very much in line with the now standard method for the 
masses involved in (\ref{eq:squarkchain}), 
this way of obtaining the gluino mass is new. 
At present two other methods are proposed. 
One is an approximate technique, used for example in \cite{Gjelsten-atlas}. 
If, in the three-body decay $\NT\to\NO ll$, we have $\mll$ at its
maximum value, $\mll=\maxmll$, then $\NO$ is at rest in the rest frame
of $\NT$.  Assuming $\mNO$
known, the four-vector of $\NT$ can then be fully reconstructed. Next,
adding the two quark four-momenta provides us with the complete
four-momentum of the gluino (and squark). This allows the mass of the
gluino (and the squark) to be found in the more traditional way of
plotting the mass peak.  Apart from the normal experimental
uncertainties this method is exact.  
However, for the decay chain (\ref{eq:gluinochain}), it is generally 
not the case that $\NO$ is at rest in the rest frame of $\NT$ 
when $\mll=\maxmll$. 
Only if $\mlR^2=\mNT\mNO$ is this true, and the more this relation is 
violated, the less reliable the results will become. 
Furthermore, to get a sizable
sample, also events with $\mll$ somewhat below $\maxmll$ must be used, 
which increases the uncertainty of the resulting masses. 
So, to apply this method to (\ref{eq:gluinochain}) the following three 
effects must be controlled:
i) the knowledge of $\mNO$ --- this can be assumed known from the 
standard endpoint analysis of the squark chain, but the appropriate 
uncertainties must be included, 
ii) the systematics from the violation of $\mlR^2=\mNT\mNO$, and  
iii) the systematics from the inclusion of events with $\mll<\maxmll$. 
A rigorous treatment of all these effects is still wanting.

The other method to obtain the gluino mass from (\ref{eq:gluinochain}) 
goes under the name of the `Mass relation method',  
and is described in \cite{Nojiri:2004}.
In a first version the masses of $\NO$, $\lR$ and $\NT$ are 
assumed known, e.g.\ 
from the squark-endpoint analysis. For a given gluino-chain event 
we then have six unknowns: $\mgl$, $\mqL$ and the four-momentum of $\NO$. 
On-shell mass conditions for the five sparticles provide five equations. 
For {\it one} such event we are one condition short of determining the system. 
[If also $\mqL$ is assumed known, the system is solvable.]
A second event of the same type will again give 5 equations for 6 unknowns. 
Two of the unknowns, $\mgl$ and $\mqL$ are however the same for the 
two events. In combination one therefore has 10 equations for 10 unknowns. 
The system can be solved analytically to provide (in some cases multiple) 
masses for the gluino and the squark. 
The mass relation method involves no approximations of the type described 
for the first method, and is in 
this respect at the same level as the endpoint method we will use here. 
The mass relation method 
also has the advantage that it avoids the difficulties related 
to estimating endpoints, and it can give a measurement even for 
quite a small number of events. 
On the other hand the uncertainties on the three lighter masses 
as well as possible correlations between the masses need to be 
appropriately handled. 
An extension of the method to take no input masses, 
and/or include in a consistent way measurements from an 
endpoint analysis should be feasible and would meet these requirements. 

For the endpoint method, which in the following is developed to also 
provide the gluino mass, all measurements go into the same fit, 
so errors and correlations are automatically treated correctly.

The structure of the present paper is as follows. 
In Sect.~\ref{sect:calc} we present the endpoints of the invariant
distributions associated with gluino decays. We also discuss the
ambiguity which arises from having two quarks in the final state and
plot some examples of gluino distributions with no experimental
effects. Sect.~\ref{sect:analysis} presents a simulation of the gluino
endpoints at the ATLAS detector for the benchmark SUSY scenario
SPS~1a, ending with an estimate for some of the new endpoints. Energy
scale errors are also discussed in some detail.  In
Sect.~\ref{sect:masses} values for masses and mass differences are
found from the estimates of the endpoint measurements. 
In Sect.~\ref{sect:doublederror} the impact of the statistical error 
is separated from that of 
the energy scale error 
and two somewhat more conservative error scenarios are studied. 
The effect of adding a Linear Collider measurement is
investigated in Sect.~\ref{sect:LC}, while conclusions are made in
Sect.~\ref{sect:conclusion}. 
A new method for calculating $\maxmqFll$ 
is developed and generalised to the gluino case $\maxmqqll$ 
in App.~\ref{app:mqqll}. 
Finally, the complicated gluino endpoint 
${\maxmqqlLow}$ is derived in some detail in 
App.~\ref{app:mqqlLow}.

\section{Gluino endpoints}\label{sect:calc}

\subsection{Endpoint calculation}
Using the method detailed in App.~\ref{app:mqqll} we obtain the 
maximum value of $\mqqll$ for the five-sparticle gluino chain: 
\begin{eqnarray}
\left(\maxmqqll\right)^2 
&=& \left\{ 
\begin{array}{llcc}
\frac{\big(\sgl-\sqL\big)\big(\sqL-\sNO\big)}{\sqL} 
& \ {\rm for } \quad 
& \rglqL > \rqLNT\rNTlR\rlRNO & \quad {\itB (1)} \\[4mm]
\frac{\big(\sgl\sNT-\sqL\sNO\big)\big(\sqL-\sNT\big)}{\sqL\sNT} 
& \ {\rm for } \quad
& \rqLNT > \rNTlR\rlRNO\rglqL & \quad {\itB (2)}\\[4mm]
\frac{\big(\sgl\slR-\sNT\sNO\big)\big(\sNT-\slR\big)}{\sNT\slR} 
& \ {\rm for } \quad
& \rNTlR > \rlRNO\rglqL\rqLNT & \quad {\itB (3)} \\[4mm]
\frac{\big(\sgl-\slR\big)\big(\slR-\sNO\big)}{\slR} 
& \ {\rm for } \quad
& \rlRNO > \rglqL\rqLNT\rNTlR & \quad {\itB (4)} \\[4mm]
\big(\mgl-\mNO\big)^2 
& \ \lefteqn{\rm otherwise} && \quad {\itB (5)} 
\end{array} 
\right\}  \label{eq:edge-qqll} 
\end{eqnarray}
Notice the systematic form of the defining inequalities and the 
resemblence to $\maxmqFll$, given in \REFqFll. 
For the gluino endpoint we have four
`dominance regions', each defined in terms of a nearest-neighbour mass
ratio dominating the product of the three others. As before,
simplifications can be made to these inequalities, but with the
undesired consequence of obscuring the neat and systematic structure.

In principle, a measurement of $\maxmqqll$ would, in combination 
with the squark endpoints, suffice to provide the gluino mass. 
The long decay chain does however allow for more 
distributions to be constructed, from which endpoints can be 
extracted and compared with analytic expressions. 
These other available endpoints are useful both 
as over-constraining measurements and as consistency tests.

The gluino chain has seven primary endpoints which involve $\mgl$.  
The two-particle primary endpoints $\maxmqq$, $\maxmqNlN$, $\maxmqNlF$ are 
easily calculated from cascade decays `on a line'. 
There is only one realisation for each. 

For the three-particle endpoint $\maxmqqlN$, 
the particles involved are neigbours in the decay chain, 
giving a situation similar to $\maxmqFll$. 
The solution can be found from \REFqFll\ 
via the substitutions 
\begin{equation}
(\mqL,\mNT,\mlR,\mNO)\to(\mgl,\mqL,\mNT,\mlR)
\end{equation}
in both the endpoint expressions and the inequalities. 

The two other three-particle endpoints $\maxmqNll$ and  $\maxmqqlF$ are 
more difficult to find because the particles involved are not all nearest 
neighbours in the decay chain. 
However, it is possible to transform the problem into one which involves 
only nearest neighbours. 
Let us first consider $\maxmqNll$. 
The initial point of difficulty is the orientation of the $\qL$ decay, 
where the unused $\qF$ is emitted. 
Will the orientation which gives the largest $\mqNll$ depend 
on the mass scenario? 
To answer this question, boost to the rest frame of $\NT$. 
Whatever the details of the two steps prior to the creation of $\NT$, 
at this stage the only quantity of relevance to $\mqNll$ 
is $|\vvec{p}_{\qN}|$. The larger the momentum is, the larger $\mqNll$ 
can become. Thus, independent of masses, for maximum values of $\mqNll$, 
the orientation of the $\qL$ decay is always the one that maximises 
$|\vvec{p}_{\qN}|$ in the rest frame of $\NT$. 
This means sending $\NT$ in the opposite direction of $\qN$, and
the configuration which gives $\maxmqNll$ therefore has 
$\vvec{p}_{\qN}$ parallel to $\vvec{p}_{\qF}$.

With this point settled, the situation can be transformed into a 
nearest-neighbour decay, for which we know the solution. 
To do this we introduce a pseudo-particle:
as seen from the rest frame of $\NT$ the momentum of $\qN$ corresponds 
to the decay of 
a pseudo-particle $\YY\to\NT\qN$ with mass $m_{\YY}^2=\mgl^2-\mqL^2+\mNT^2$. 
In such a picture all particles are nearest neighbours, and the solution 
\REFqFll\ 
applies with the appropriate substitution
\begin{equation}
\mqL\to m_{\YY}=\sqrt{\mgl^2-\mqL^2+\mNT^2}
\end{equation}

For $\maxmqqlF$ the orientation of the $\NT$ decay is fixed to maximise 
$|\vvec{p}_{\lF}|$ in the rest frame of $\NT$, analogous to the argument 
above. Another pseudo-particle is then defined at the end of the decay chain, 
$\NT\to \YY\lF$ where $m_{\YY}=\mNO\mNT/\mlR$. Again, the problem has been 
reformulated in terms of nearest neighbours, so solution 
\REFqFll\ 
applies. The appropriate substitutions are 
\begin{equation}
(\mqL,\mNT,\mlR,\mNO) \to (\mgl,\mqL,\mNT,\mNO\mNT/\mlR)
\end{equation}

As was discussed for the squark chain, primary distributions 
based on distinguishing the two leptons can in practice not be 
constructed. Instead of the distributions $\mqNlN$, $\mqNlF$, $\mqqlN$ 
and $\mqqlF$ one can construct $\mqNlHigh$, $\mqNlLow$, $\mqqlHigh$ and 
$\mqqlLow$. 
While the endpoint of a `$\high$' distribution is simply the maximum 
of the two primary endpoints for the given mass scenario, 
further considerations are needed for the `$\low$' distributions. 
For $\mqNlLow$ these considerations are quite manageable,  
but for the three-particle distribution $\mqqlLow$ they become rather involved.
The general solution for the latter is given in App.~\ref{app:mqqlLow}. 

The endpoint expressions are summarised below. 
Together with Eq.~(\ref{eq:edge-qqll}), the six first ones, 
Eqs.~(\ref{eq:edge-qq})--(\ref{eq:edge-qNlHigh}),  
correspond to the distributions which 
are usable in that they do not rely on distinguishing the two leptons:
\begin{eqnarray}
\left(\maxmqq\right)^2 
&=& \big(\sgl-\sqL\big)\big(\sqL-\sNT\big)/\sqL 
\label{eq:edge-qq} \\[4mm]
\mqqlLow^\max
&:& \mbox{see App.~\ref{app:mqqlLow}}
\label{eq:edge-qqlLow1} \\[4mm]
\mqqlHigh^\max
&=& \max(\maxmqqlN,\maxmqqlF)
\label{eq:edge-qqlHigh} \\[4mm]
\maxmqNll
&:& \mbox{\REFqFll\ with}\ 
\mqL\to\sqrt{\sgl-\sqL+\sNT}
\label{eq:edge-qNll} \\[4mm]
\mqNlLow^\max
&=& \max(\maxmqNlN,m_{\qN l(\equal)}^\max)
\label{eq:edge-qNlLow} \\[4mm]
\mqNlHigh^\max
&=& \max(\maxmqNlN,\maxmqNlF)
\label{eq:edge-qNlHigh} \\[4mm]
\left(\maxmqNlN\right)^2 
&=& \big(\sgl-\sqL\big)\big(\sNT-\slR\big)/\sNT 
\label{eq:edge-qNlN} \\[4mm]
\left(\maxmqNlF\right)^2 
&=& \big(\sgl-\sqL\big)\big(\slR-\sNO\big)/\slR 
\label{eq:edge-qNlF} \\[4mm]
\big(m_{\qN l(\equal)}^\max\big)^2 
&=& \big(\sgl-\sqL\big)\big(\slR-\sNO\big)/(2\slR-\sNO) 
\label{eq:edge-qNlEq} \\[4mm]
\maxmqqlN
&:& \mbox{\REFqFll\ with}\ 
(\mqL,\mNT,\mlR,\mNO)\to(\mgl,\mqL,\mNT,\mlR)
\label{eq:edge-qqlN} \\[4mm]
%
%
\maxmqqlF
&:& \mbox{\REFqFll\ with}\
(\mqL,\mNT,\mlR,\mNO)\to(\mgl,\mqL,\mNT,\mNO\mNT/\mlR)\nonumber\\&&
\mbox{\hspace{1.3cm}}
\label{eq:edge-qqlF} \\[-2mm]\nonumber
\end{eqnarray}

\subsection{Quark ambiguity\label{subsect:quarkambiguity}}

Adding a gluino to the cascade introduces an additional ambiguity: how
does one distinguish between the jet initiated by the gluino decay and
that initiated by the squark decay? Initially we will assume that the
two correct jets of a signal event have been selected, although we 
do not know which is which. (The difficulties associated with this 
assumption will be commented on later.) In this case, one
can seek to distinguish the two jets kinematically, by, for
example, their transverse momenta. If $\sgl-\sqL$ is very
different from $\sqL-\sNT$, the two quarks will be emitted with
noticeably different energies, and a $p_T$ comparison will, with a
certain purity, serve to distinguish between the jets.  However, one
cannot, a priori, know which jet is to be assigned to $\qN$ or $\qF$,
i.e.\ whether it is $\qN$ or $\qF$ which gives rise to the hardest
jet, and one must instead turn to the differences in production rates.
Only when a gluino sits on top of the chain is $\qN$ produced, and
often, the ratio of directly versus indirectly produced squarks is
sufficient to determine the average behaviour of $\qF$.  This then allows
statements on the average behaviour of $\qN$ and $\qF$ to be made,
which may be used to construct samples where the quark identity is
required, Eqs.~(\ref{eq:edge-qNll})--(\ref{eq:edge-qNlHigh}), as well
as the squark distributions $\mqFll$, $\mqFlLow$ and $\mqFlHigh$.
Such a separation of $\qN$ and $\qF$ will always introduce impurities;
whether these are manageable or not will depend on the given SUSY
scenario.

To be fully general one might consider constructing new secondary
distributions in which there is no need to distinguish the jets.  The
distributions $\mll$, $\mqq$, $\mqqll$, $\mqqlLow$ and $\mqqlHigh$
would remain unaffected, but instead of $\mqFll$ and $\mqNll$, one should
construct new distributions $m_{qll(\high)}$ and
$m_{qll(\low)}$.  The endpoint expression of the former is readily
available, but not of the latter.  The complexity of the calculation
and the resulting expression for $\maxmqllLow$ can be expected to be
similar to that of $\maxmqqlLow$ shown in App.~\ref{app:mqqlLow}.
Furthermore, for the four $\mql$ distributions one can collect four
new distributions based on the order of the four values per
event. Only the endpoint of the $\mqlHigh$ distribution is readily
available, while the other three endpoints become very difficult to
calculate.  A programme to calculate the new endpoints where quarks
are not distinguished, has so far not been started.

The above discussion shows that as a second jet is added, the 
complexity of the situation increases significantly. In return many 
more endpoints become available, allowing for thorough consistency checks. 
In Sect.~\ref{sect:analysis} we will return to these issues in the 
context of a simulation at the standard SPS~1a point.

\FIGURE[ht]{
\epsfxsize 12.7cm 
\epsfbox{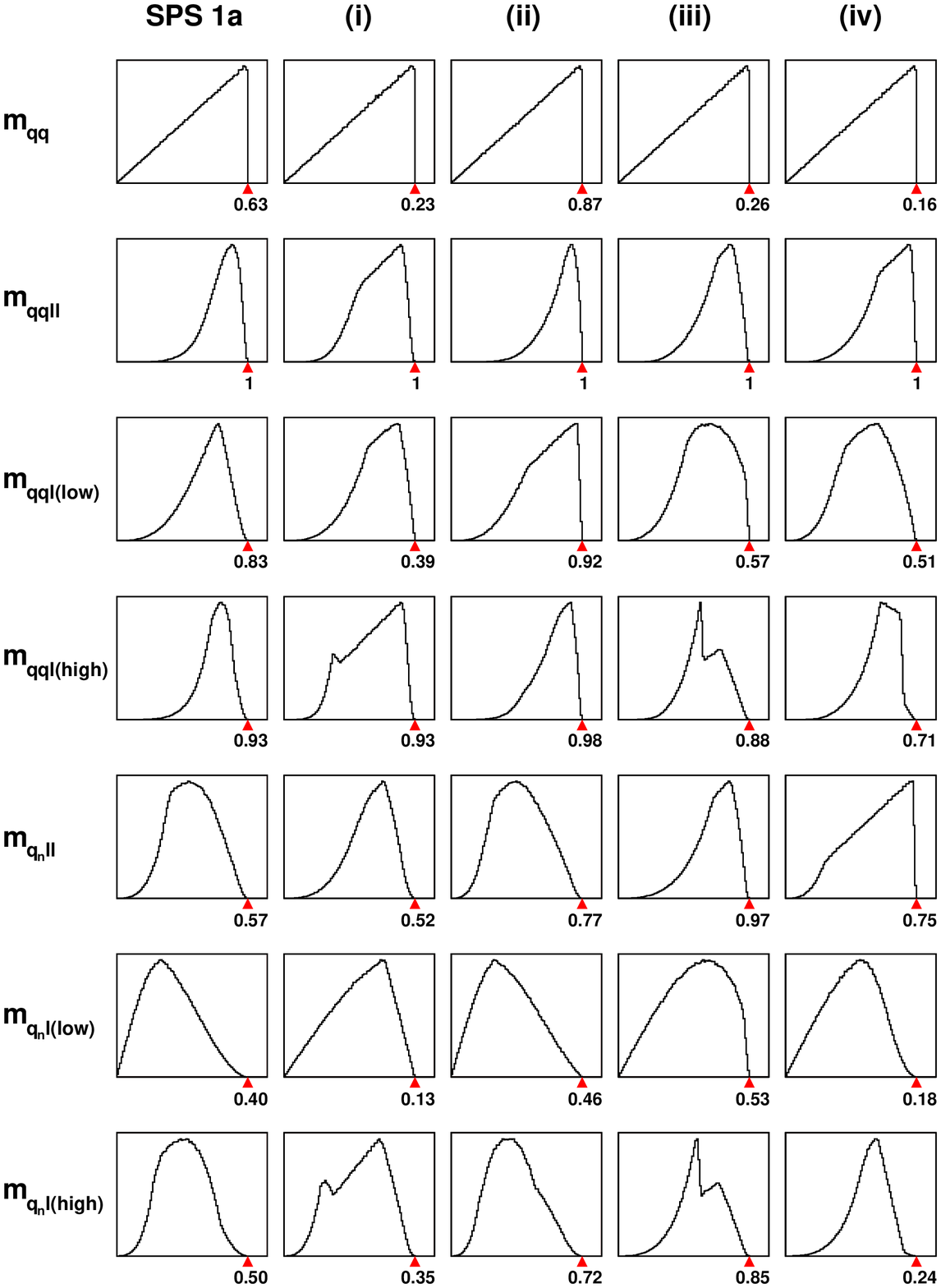}
\caption{Theory distributions for \SPSOa\ and four other mass scenarios, 
showing some of the shape variety. Only gluino distributions are shown.
Kinematic endpoints are marked with a triangle and given in units of 
$\maxmqqll$, the largest of the endpoints. 
(All panels start at zero invariant mass.) 
The scenarios (i)--(iv) are defined by the following sets of 
%
%
mass ratios $\vvec{r}\equiv(\mgl/\mqL,\mqL/\mNT,\mNT/\mlR,\mlR/\mNO)$: 
$\vvec{r}_{\rm(i)}=(1.03,1.30,1.07,6.01)$, 
\mbox{$\vvec{r}_{\rm(ii)}=(1.79,3.01,1.15,1.59)$}, 
$\vvec{r}_{\rm(iii)}=(1.45,1.05,1.27,5.08)$ and 
$\vvec{r}_{\rm(iv)}=(1.03,1.29,2.02,3.60)$.
\vspace{-0.0cm}
\label{fig:TheoryDistr}}}

\subsection{Theory distributions}
In order for the endpoint method to be useful, not only must the analytic
expressions for the endpoints be available, it must also be possible to
determine these endpoints from the experimental mass distributions.  A first
criterion for this is that the edges of the distributions `point'
unambiguously towards the exact endpoints. If the shape of a distribution is
sufficiently concave at high values, the endpoint will most likely be
underestimated and large systematic uncertainties must be added.  In general
the shapes of the mass distributions vary if the sparticle masses are
varied. To investigate the range of possible shapes for each distribution, the
full cascade was generated for approximately 500 unique mass scenarios, 
and the resulting theoretical mass distributions studied visually. 

A representative selection of mass scenarios, showing some of the
shape variety of the gluino distributions, is given in
Fig.~\ref{fig:TheoryDistr}.
In generating these decays, quarks and leptons are assumed to be
massless, which in the worst case ($b$-quark) leads to a perfectly
acceptable error of $\mathcal{O}$(MeV).  Furthermore, only decay phase
space has been used, without the associated matrix elements. In
principle, this ignores possible spin correlations between 
leptons and/or quarks. 
However, since the distributions used here are 
summed over lepton charge, no spin effects are 
expected~\cite{Richardson:2001df}. A similar investigation for
the squark distributions was performed in \cite{Gjelsten:2004}.
See also \cite{Gjelsten-thesis}.

Since $\mqq$ is constructed from two nearest neighbours in the decay chain, 
the shape is triangular for any masses, similar to $\mll$. 
Its endpoint can thus in principle be determined quite accurately.
The shape of the $\mqqll$ distribution depends on the masses, but in
practically all scenarios the edge is well described by a linear
descent towards the theoretical endpoint. This should guarantee its
determination experimentally if the background is sufficiently low.
For most scenarios this is also true for $\mqqlLow$, although the
distribution can take on other shapes.  

The three `high' distributions, $\mqqlHigh$, $\mqFlHigh$ 
and $\mqNlHigh$, show a large variety of forms. The available shapes 
of the $\mqFlHigh$ distribution are very similar to those of the $\mqNlHigh$
distribution and are not reproduced here (see \cite{Gjelsten:2004}). A
common feature of these distributions is the danger of not noticing
the `foot', seen, for example, in $\mqqlHigh$, scenario~(iv), where
the true maximum value may be obscured by backgrounds.  

The last of the three-particle distributions, $\mqNll$, also has an
extensive variety of shapes.  The behaviour of the edge is usually
reasonable, but with a certain danger of foot-like structures, as in
scenario~(i).  Finally, the two $\mqNl$ distributions are often
unreliable in that the edges are concave. It may therefore be
difficult to get good estimates of the endpoints.

In summary, from the 500 mass scenarios of which the distributions 
in Fig.~\ref{fig:TheoryDistr} comprise a fairly representative selection, 
the following can be stated regarding their overall behaviour: 
The distributions 
$\mqq$ and $\mqqll$ point reliably towards the endpoint for any mass
scenario.  Somewhat less reliable are the endpoint estimates found
from the three-particle distributions. Lastly, even less reliable
endpoint estimates can be obtained from the two $\mqNl$ distributions.
In the next section we will see how these statements change 
for a `realistic' experimental setup.

\section{Measuring gluino endpoints for SPS 1a}\label{sect:analysis}

Here we investigate the measurement of the endpoints associated with
gluino decays for SPS 1a at ATLAS. Unless otherwise stated, the details
of the analysis are identical to those for the squark chain, as
described in Ref.~\cite{Gjelsten:2004}, and will not be reproduced here.

\subsection{Signal and backgrounds}

The SPS~1a point is defined by the mSUGRA GUT-scale parameter values
\begin{equation}
\mHalf=250~\GeV,\ \mZero=-\AZero=100~\GeV,\ \tan\beta=10,\ \mu>0
\end{equation}
evolved down to the electroweak scale by version 7.58 of 
{\tt ISAJET}~\cite{Baer:1993ae}. 
The masses of particles at the electroweak scale can be found
in~\cite{Gjelsten:2004}.  
Fig.~\ref{fig:gluinochain} shows the signal
chain with branching ratios: to the left the production
rate of gluinos is shown, followed by the branching ratios of a gluino
into left-handed squarks and both sbottom states. Since the mass
difference between $\bO$ and the other squarks is comparable to the
mass difference between the gluino and the squarks, phase space
effects become significant, resulting in a noticeably enhanced decay
rate into~$\bO$.  

\FIGURE[ht]{
\epsfxsize 14cm
\epsfbox{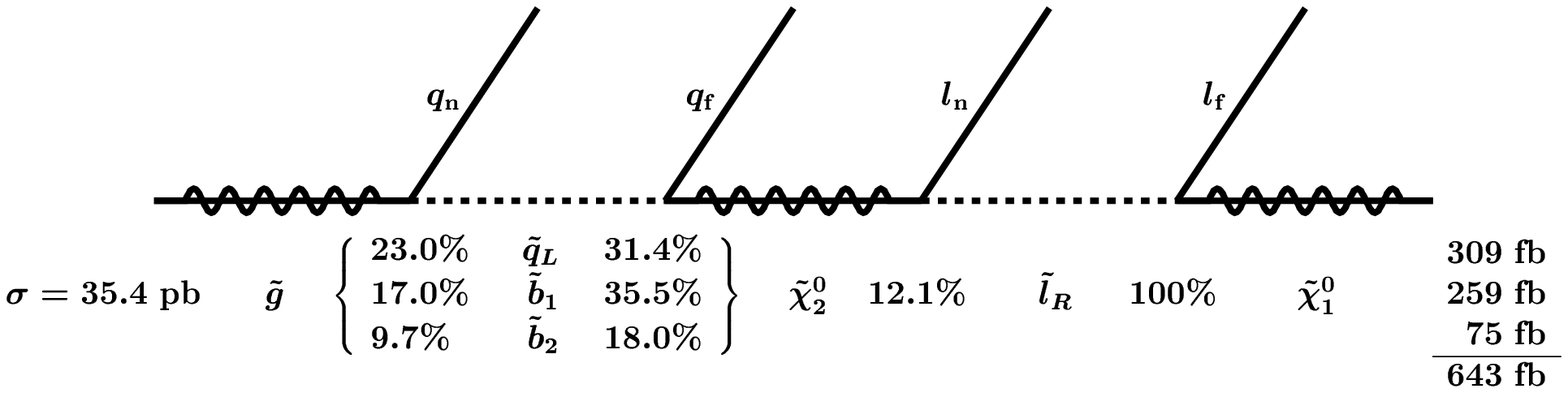}
\caption{SPS~1a cascade decay with branching ratios and cross-sections.
\label{fig:gluinochain}}}

Signal and background are generated for 300~fb$^{-1}$, which
corresponds to three years at design luminosity of $10^{34}~{\rm
cm}^{-2}{\rm s}^{-1}$.  
The low-energy parameters are passed, 
via the standard interface, to PYTHIA~6.2 \cite{PYTHIA} which 
calculates the decay widths and the LHC cross-sections by use 
of CTEQ~5L \cite{Lai:1999wy} PDF's, and generates the events. 
Finally ATLFAST~2.60 \cite{ATLFAST} performs a parametrised
fast simulation of the ATLAS detector.  The simulation setup is
identical to that used for the squark-endpoint analysis, except that
four jets are now required with $p_T^\textrm{jet}>150, 100, 50,
20$~GeV, rather than only three. Also, the definition of $M_{\rm eff}$ 
is extended to include the fourth jet.

The QCD background is cut away by the requirement of two
leptons\footnote{For convenience of notation we will for the rest of
the paper use `lepton'/`slepton' for the two first generations and
`tau'/`stau' for the third generation.}
and of considerable missing $p_T$. For the processes involving $Z$ and
$W$ the requirement of high hadronic activity together with the
missing $p_T$ removes nearly all events. After these rather hard cuts,
the Standard Model background consists of approximately 95\% $t\bar
t$. 
Most likely the Standard Model background is significantly 
underestimated, as the parton shower approach of {PYTHIA} does not 
generate at a realistic rate the hard jets which are required in  
the selection cuts. 
However, because of the large SUSY cross-section for this scenario, 
the main background will come from other SUSY processes. 
The possible underestimation of the Standard Model background is 
therefore not a danger for the current analysis.

It is convenient to divide the background into three different parts, 
`lepton-uncorrelated', `lepton-correlated' and `combinatorial'. 
The two first consist of events which do not contain the signal chain. 
In the {\it lepton-uncorrelated} background two leptons are produced,
but in different parts of the decay and therefore independently. With
the assumption of lepton universality and neglecting the mass
difference between electrons and muons, which is very reasonable for
the energies involved, this background can be removed by subtracting
the distribution for different-flavour leptons, as was done in
Ref.~\cite{Gjelsten:2004}.

For the {\it lepton-correlated} background, 
the leptons are always same-flavour, 
so no \linebreak[4]
different-flavour sample is available to show its distribution. 
These events typically come from the decay of $Z$, in which case they 
can to some degree be controlled, or from sleptonic decay of neutralinos. 
In the latter case no particular signature, e.g.\ in the $\mll$ distribution, 
is available to discriminate this background from the signal. 
In particular much of this background will come from $\NT$'s decaying 
sleptonically, but which are not part of our signal chain. 
For the analysis of the squark chain, as done in \cite{Gjelsten:2004}, 
one important criterion is that the $\NT$'s 
produced from squarks make up a good fraction of the total $\NT$ production. 
Events with $\NT$'s not originating from the relevant squarks usually still 
contain jets and constitute a large part of the background. 
As the signal chain grows longer, this background increases relative to 
the signal. In the case of the whole gluino chain 
what is important is that the $\NT$'s with a gluino grandparent 
are not severely outnumbered by the total $\NT$ production, for the events
selected.  This is a stricter criterion, and we will find that the flavour 
of the intermediate squark, whether it is $\qL$ or $\sb$, will be crucial 
for the isolation of the signal. 

The third background type, combinatorial background, is a result of our 
inability to know which of the jets correspond to the quark sent out from 
the gluino ($\qN$) and from the squark ($\qF$). 
In most analyses one assumes that the quark from a squark decay 
can on average be distinguished from the quark from a gluino 
decay on the basis of hardness. In most mSUGRA scenarios this 
is viable, as $\NT$ is so much lighter than the squark. 
Since two squarks are expected in nearly all events, 
assuming here $\mgl>\msq$, 
one can attribute the two hardest jets to the decay of the two squarks. 
Remaining jets can then be attributed to gluino decays, other quarks in the 
event, e.g.\ from the decay of $t$ or $W$, and/or the underlying event. 
The exact jet selection algorithms used in this study are based on these 
assumptions, and will be detailed in the subsections below.

In more general SUSY scenarios it need not be the case that $\qF$ is 
harder than $\qN$. In general, given an unknown SUSY scenario, the appropriate 
jet selection procedure must be the result of careful study. 
For this also the mass hierarchy of the gluino and the squarks must be 
established. 
A systematic study of how such information can be obtained is lacking 
but would be very valuable.

\subsection{Non-$b$-tagged distributions}

\FIGURE[ht]{
\let\picnaturalsize=N
\def\picsize{7.1cm}
\ifx\nopictures Y\else{
\let\epsfloaded=Y
\vspace*{-1mm}
\centerline{{\ifx\picnaturalsize N\epsfxsize \picsize\fi
\epsfbox{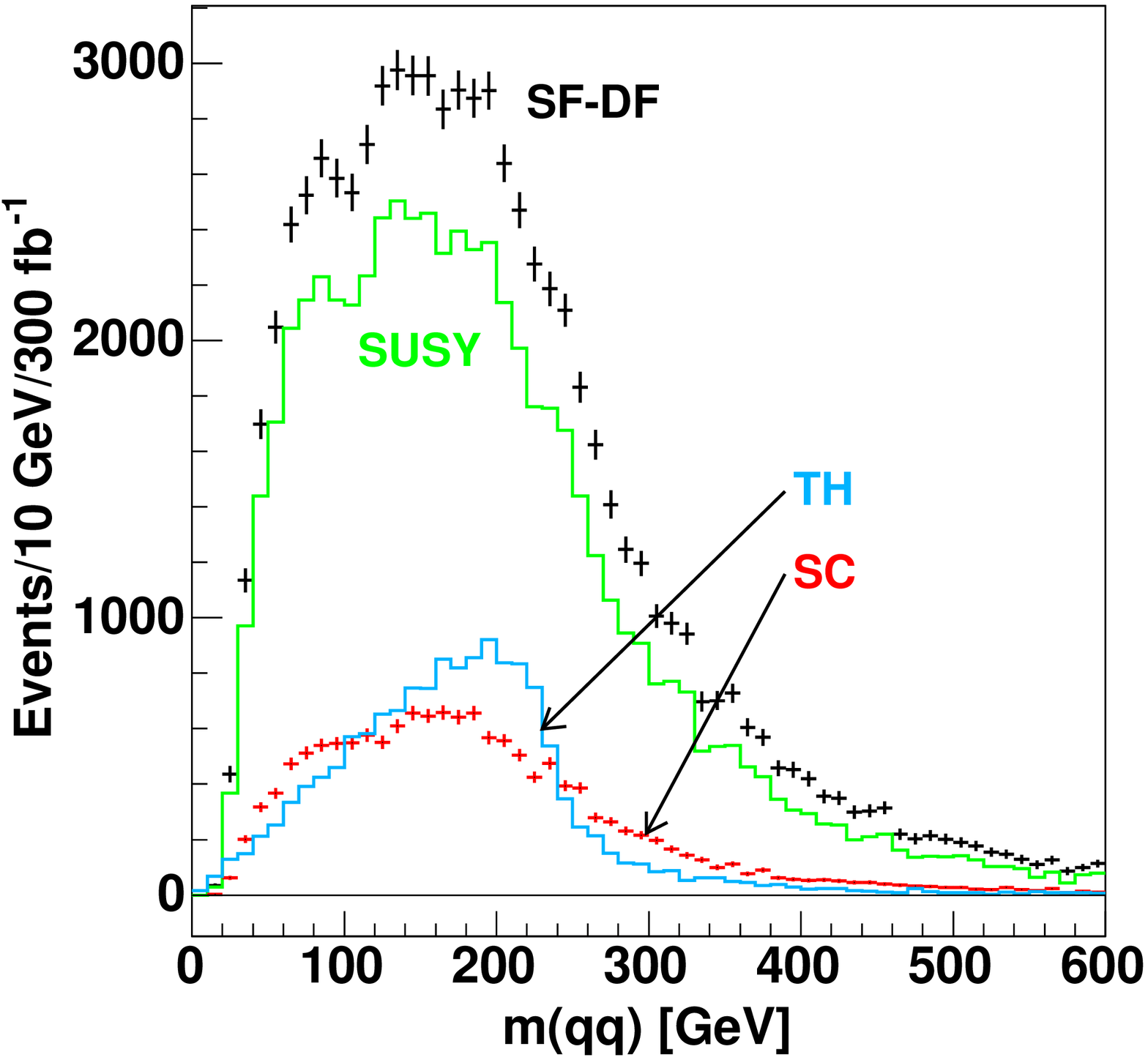}
            {\ifx\picnaturalsize N\epsfxsize \picsize\fi
\epsfbox{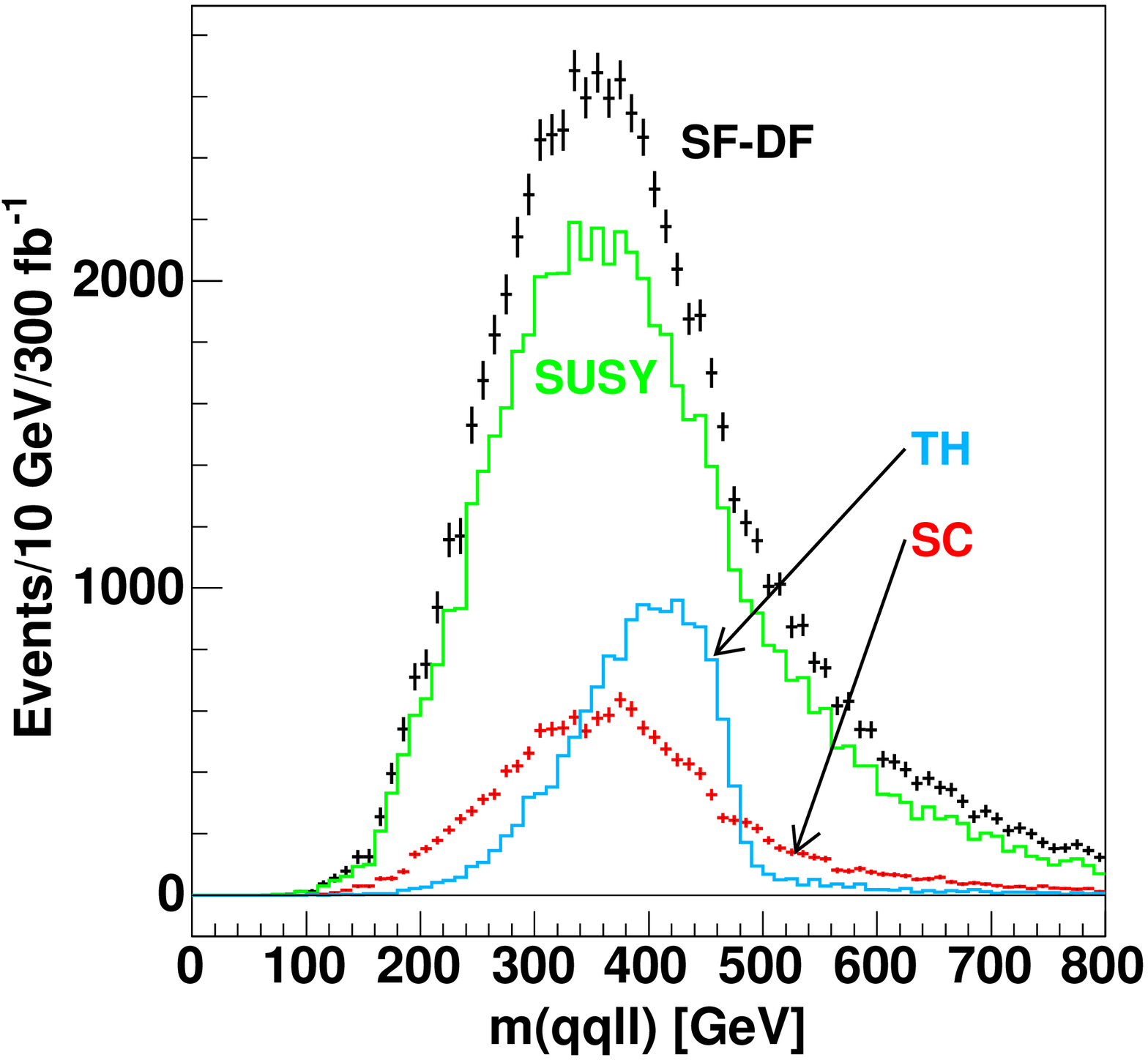}} 
} }
}\fi
\vspace*{-6mm}
\caption{Invariant mass distributions for non-$b$-tagged events. 
Endpoints not detectable. See the text for details. 
\label{fig:mqq}}}

In Fig.~\ref{fig:mqq} two of the non-$b$-tagged distributions, 
$\mqq$ and $\mqqll$, are shown. 
%
In black with error bars the different-flavour-subtracted distribution 
(`SF-DF') is shown. 
Solid green shows the SUSY background (`SUSY'). Its shape is given by 
the lepton-correlated part, but the lepton-uncorrelated part is also
significant and increases the random fluctuations. 
The Standard Model background is negligible and is not shown separately. 
The blue curve shows the parton-level distribution 
of the selected events (`TH'), while the red points with error bars 
show the part of the different-flavour-subtracted sample which contains 
the correct signal chain (`SC'), 
also referred to as the `signal-chain distribution'.
Any discrepancy between these two distributions is mostly due to 
combinatorial background, i.e.\ from picking the wrong jets. 

For these distributions $\qF$ was assumed to be one of the two hardest 
jets while $\qN$ was selected among number three or four in $p_T$-hardness.
This is in line with the mass assumptions usually valid in mSUGRA scenarios, 
as described above. 
The jet selection used for the distributions plotted was, on an event-by-event 
basis, the one out of four possible combinations which gave the smallest 
$\mqqll$ value. 
In addition we required that neither of the involved jets was $b$-tagged. 
(The $b$-tagging simulation is described in \cite{Gjelsten:2004} and 
is rather crude, although with a reasonable overall behaviour.)

It is clear that the positions of the endpoints, which are seen in the 
parton-level distributions (`TH'), are not easily detectable from the 
different-flavour-subtracted sample. 
This is mainly due to the size of the lepton-correlated part of the 
SUSY background 
(`SUSY'), which makes up $\sim80$\% of the total sample. 
Structures in the signal part of the sample are therefore not easily
identified.
This result can be anticipated from Table~3 of Ref.~\cite{Gjelsten:2004},
which shows that only $8.2/32.8=25\%$ of $\qL$'s originate from a gluino. 
The ratio of gluino-induced $\NT$'s becomes therefore quite small. 

Another difficulty is the combinatorial background. 
The signal-chain distributions (`SC') 
do not at all point to the nominal endpoints at 242~GeV and 490~GeV 
(for $\uL$). 
This is because the jet pair selected is often not the correct one. 

Other jet selection algorithms of this simple type have been tried, 
but none allows any edge structure caused by the kinematics of the decay 
chain to be recognised. 
This is also true for the other five gluino distributions (not shown). 
One must therefore conclude that the endpoints of the non-$b$-tagged gluino 
distributions are not experimentally obtainable 
by looking at one distribution at a time. 
If instead correlations between mass distributions were investigated 
it might be possible to identify endpoint-related edge structures.

\subsection{$b$-tagged distributions}

The distributions of the $b$-tagged samples are shown in 
Figs.~\ref{fig:mbb_1}--\ref{fig:mbb_2}. 
The curves follow the colour code of Fig.~\ref{fig:mqq}, 
except the Standard Model background (`SM') is also shown 
(green points with error bars). 
For the $b$-tagged distributions the different-flavour distribution 
is $\sim40$\% of the same-flavour distribution, which is somewhat 
larger than for the non-$b$-tagged distribution where the ratio is 
$\sim25$\%. 
In both cases the different-flavour distribution favours 
lower mass values and does not interfere very much with the edge structure. 

Contrary to the previous case, the $b$-tagged distributions have clear 
edge structures which provide values for the endpoints. 
The main reason for this is that the different-flavour-subtracted 
SUSY background (solid green) 
now makes up a manageable 35\%, to be compared with 80\% 
for the non-$b$-tagged sample. 
This reduction is due to the fact that {\it the majority of $\sb$'s are produced 
indirectly from gluino decay} [because of the low $b$-content of the proton] in combination with 
a jet selection requirement of exactly two $b$-tagged jets. 
Although the total production of $\NT$ is 6--7 times larger 
(see Fig.~\ref{fig:gluinochain})
than the 
production which starts out from $\gl\to\sb b$, two $b$-jets are also needed, 
which reduces considerably the number of selected background events. 
(For background events the $b$'s are usually produced in 
the other chain, typically from the same
$\gl\to\sb b$ which then does not continue sleptonically, 
or from $\gl\to\tO t$ in either of the chains, 
which also produces multiple $b$'s.)
Since the rate of $b$-jets is considerably smaller than the rate of 
light jets, also in SUSY events, we are likely not to have additional 
$b$-jets in a signal event. The combinatorial background is therefore 
small, as is clear from the good correspondence between the parton-level 
distribution (`TH') and the signal-chain distribution (`SC').

The specific jet-selection used here is in line with the previous assumption 
that $\bF$ is harder than $\bN$. The first is searched for among the two 
$p_T$-hardest, the second among number three and four. Only events which have 
one $b$-jet among the two hardest and one among the two next were selected. 
(In a more realistic study where emphasis is put on issues like 
fitting techniques, impact on the distributions from the precuts etc., 
it would be natural to investigate the effect of also including events which 
have their two $b$-tagged jets as number 1 and 2 or as number 3 and 4.)
%

\FIGURE[ht]{
\let\picnaturalsize=N
\def\picsize{7.1cm}
\ifx\nopictures Y\else{
\let\epsfloaded=Y
\vspace*{-1mm}
\centerline{{\ifx\picnaturalsize N\epsfxsize \picsize\fi
\epsfbox{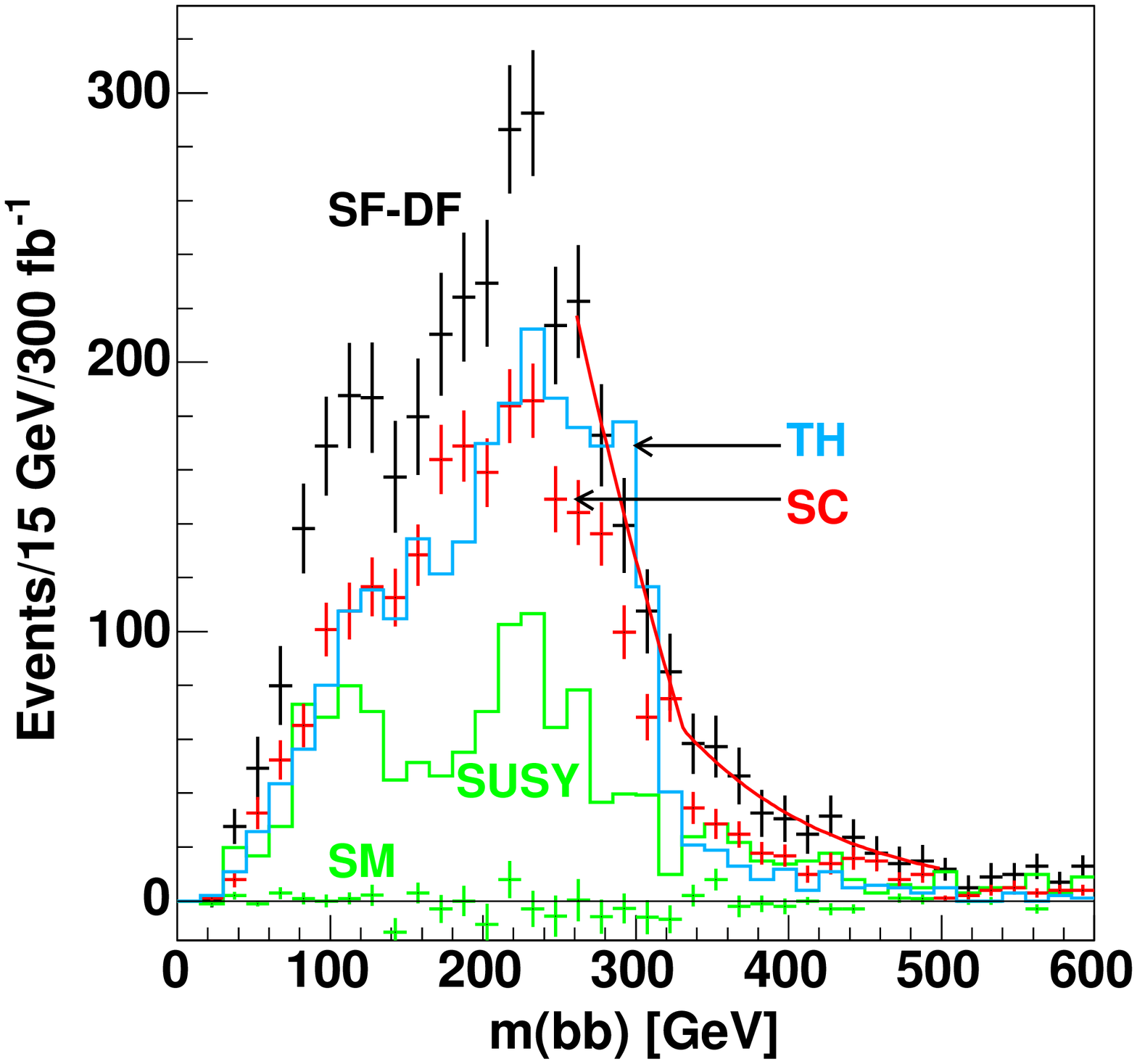}
            {\ifx\picnaturalsize N\epsfxsize \picsize\fi
\epsfbox{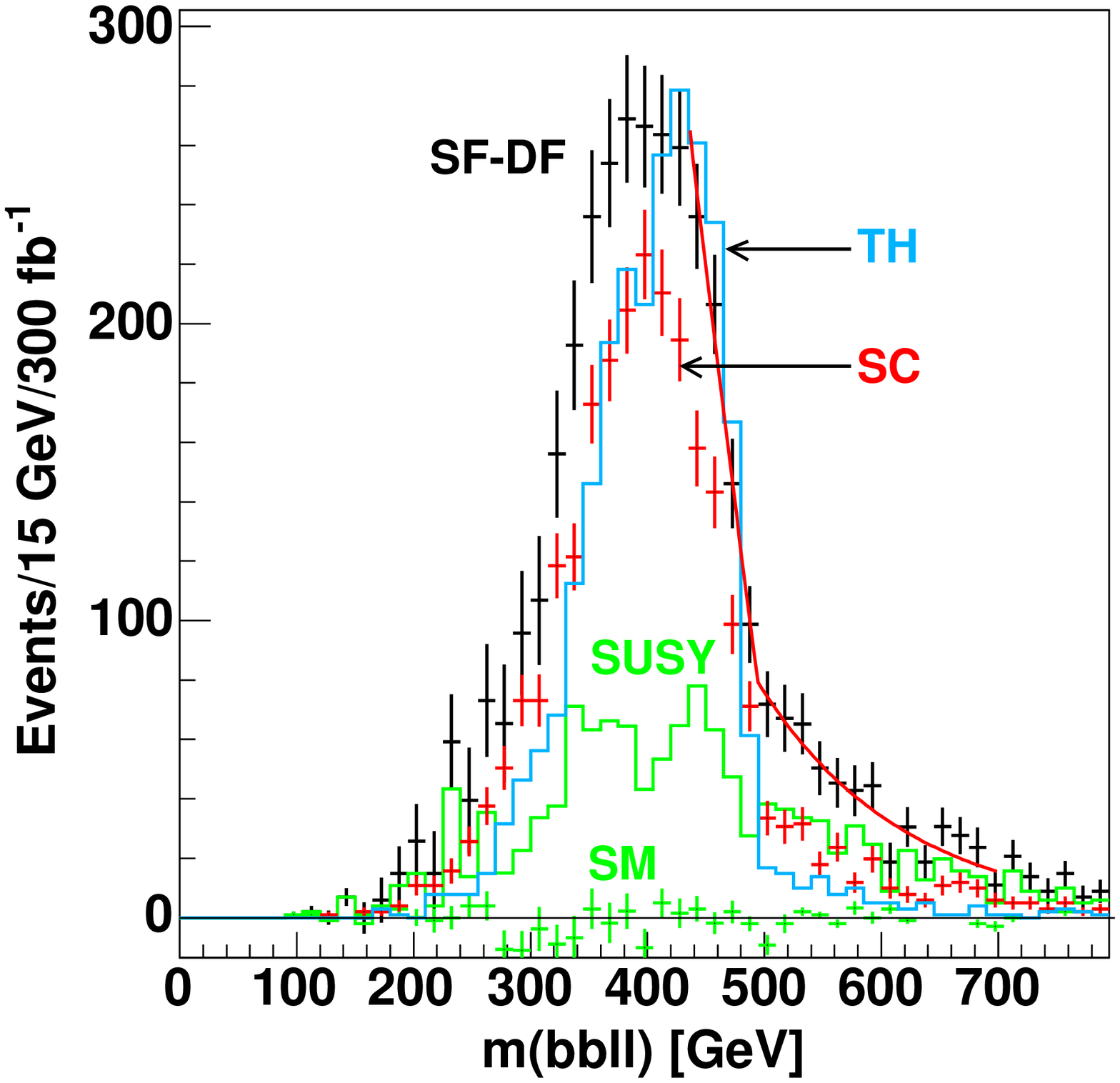}
} } }
\centerline{{\ifx\picnaturalsize N\epsfxsize \picsize\fi
\epsfbox{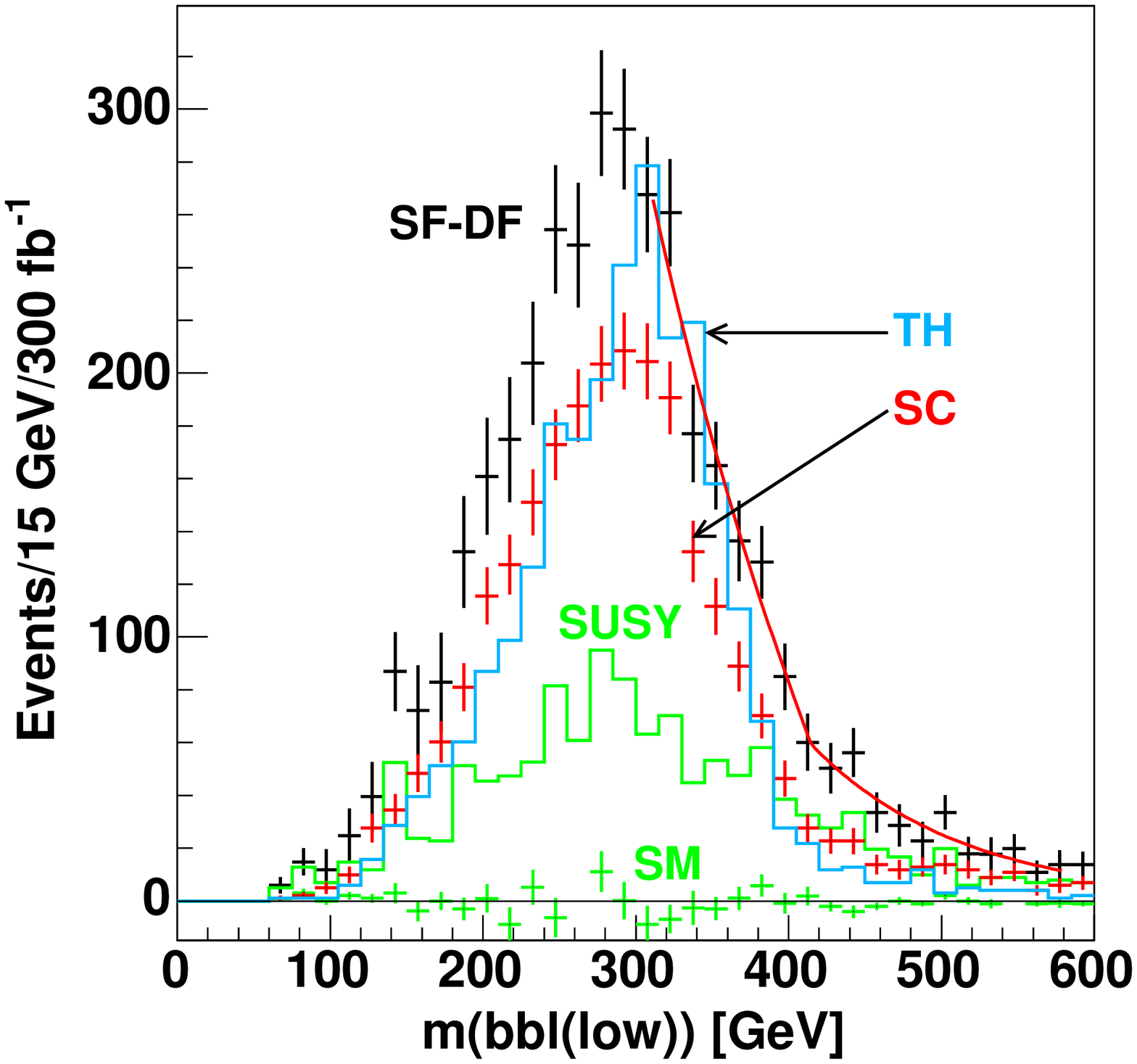}
            {\ifx\picnaturalsize N\epsfxsize \picsize\fi
\epsfbox{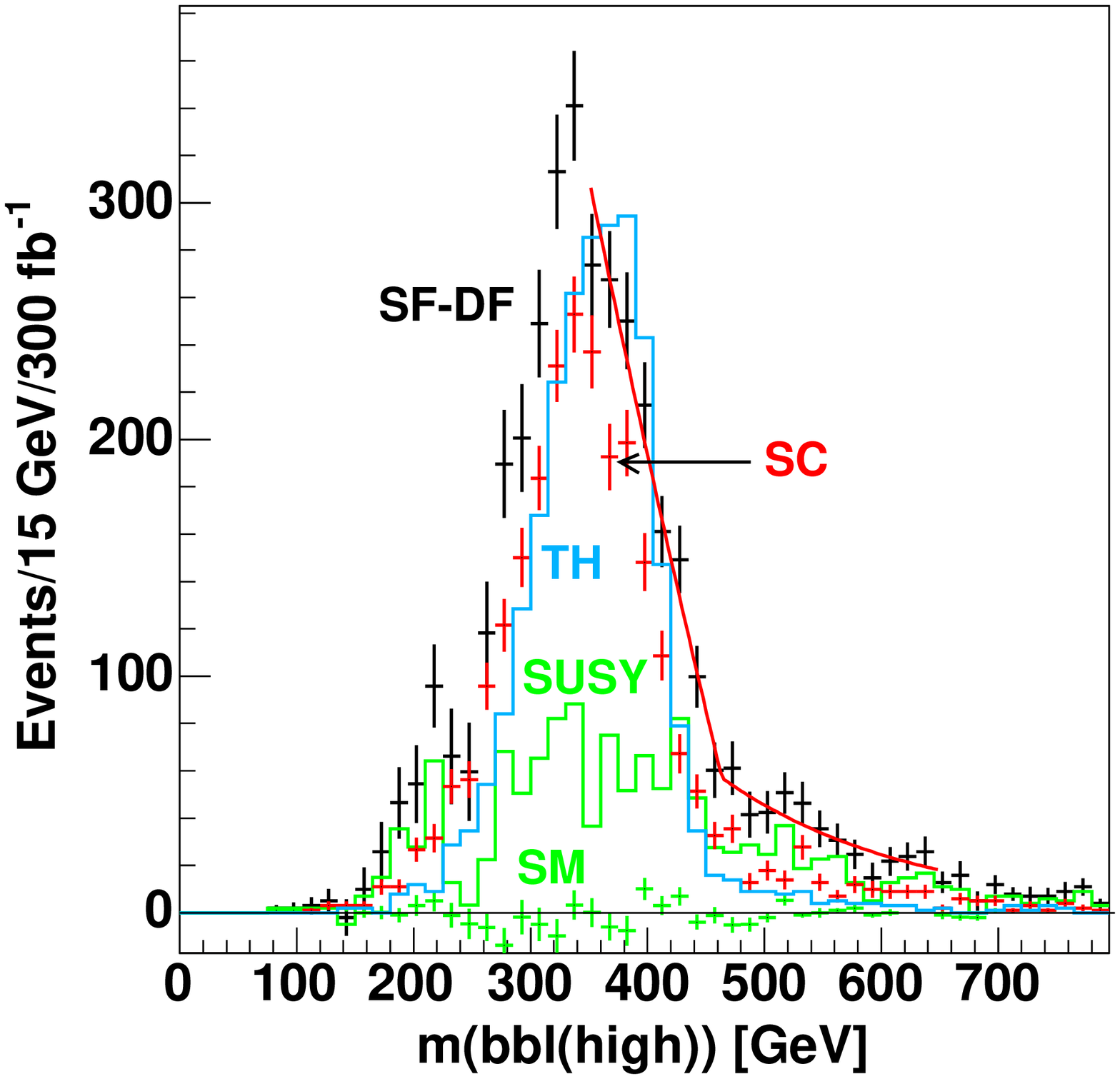}
} } }
}\fi
\vspace*{-6mm}
\caption{Four good $b$-tagged invariant mass distributions. See the 
text for details. \label{fig:mbb_1}}}

\FIGURE[ht]{
\let\picnaturalsize=N
\def\picsize{7.1cm}
\ifx\nopictures Y\else{
\let\epsfloaded=Y
\vspace*{-1mm}
\centerline{{\ifx\picnaturalsize N\epsfxsize \picsize\fi
\epsfbox{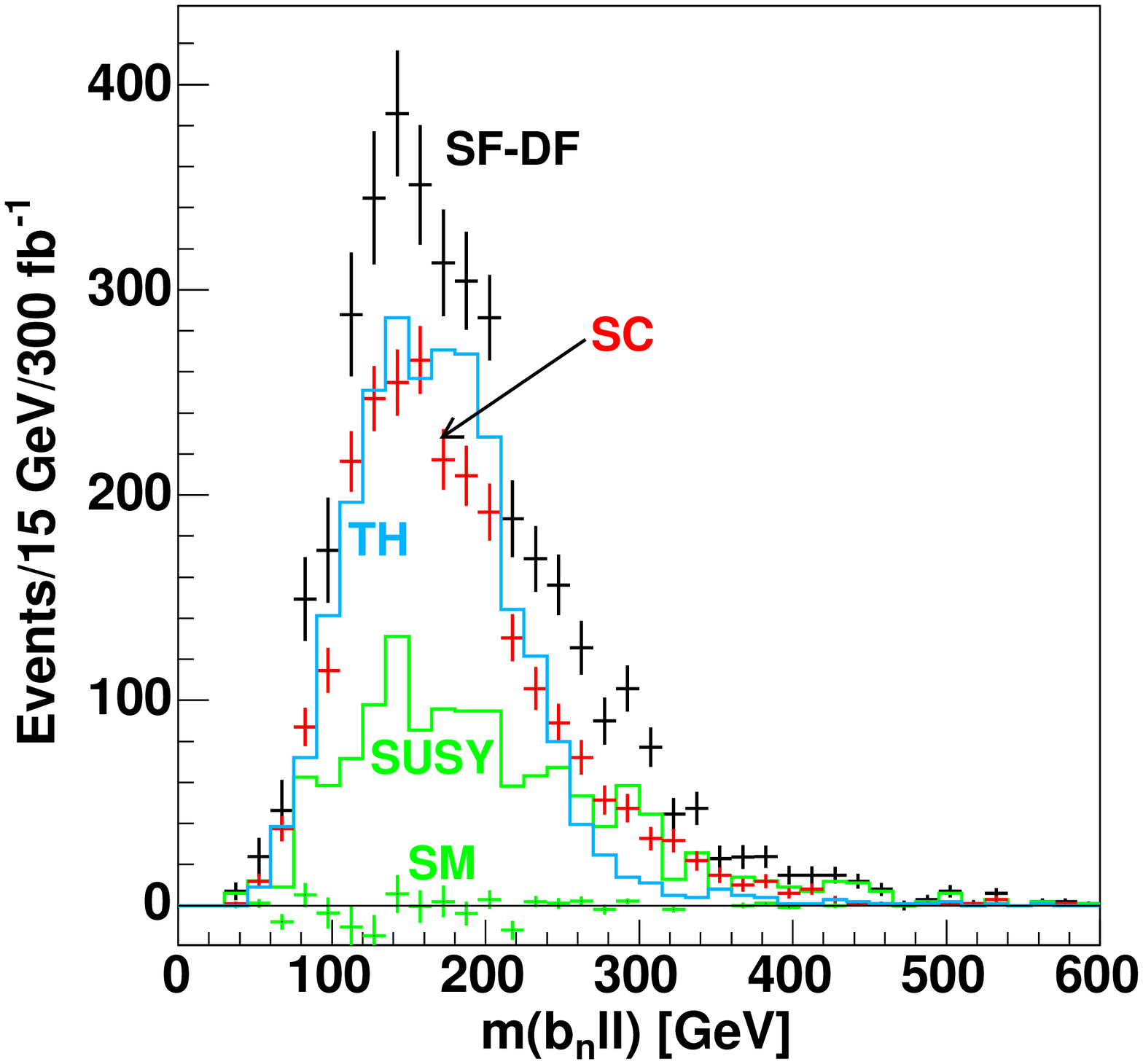}
            {\ifx\picnaturalsize N\epsfxsize \picsize\fi
\epsfbox{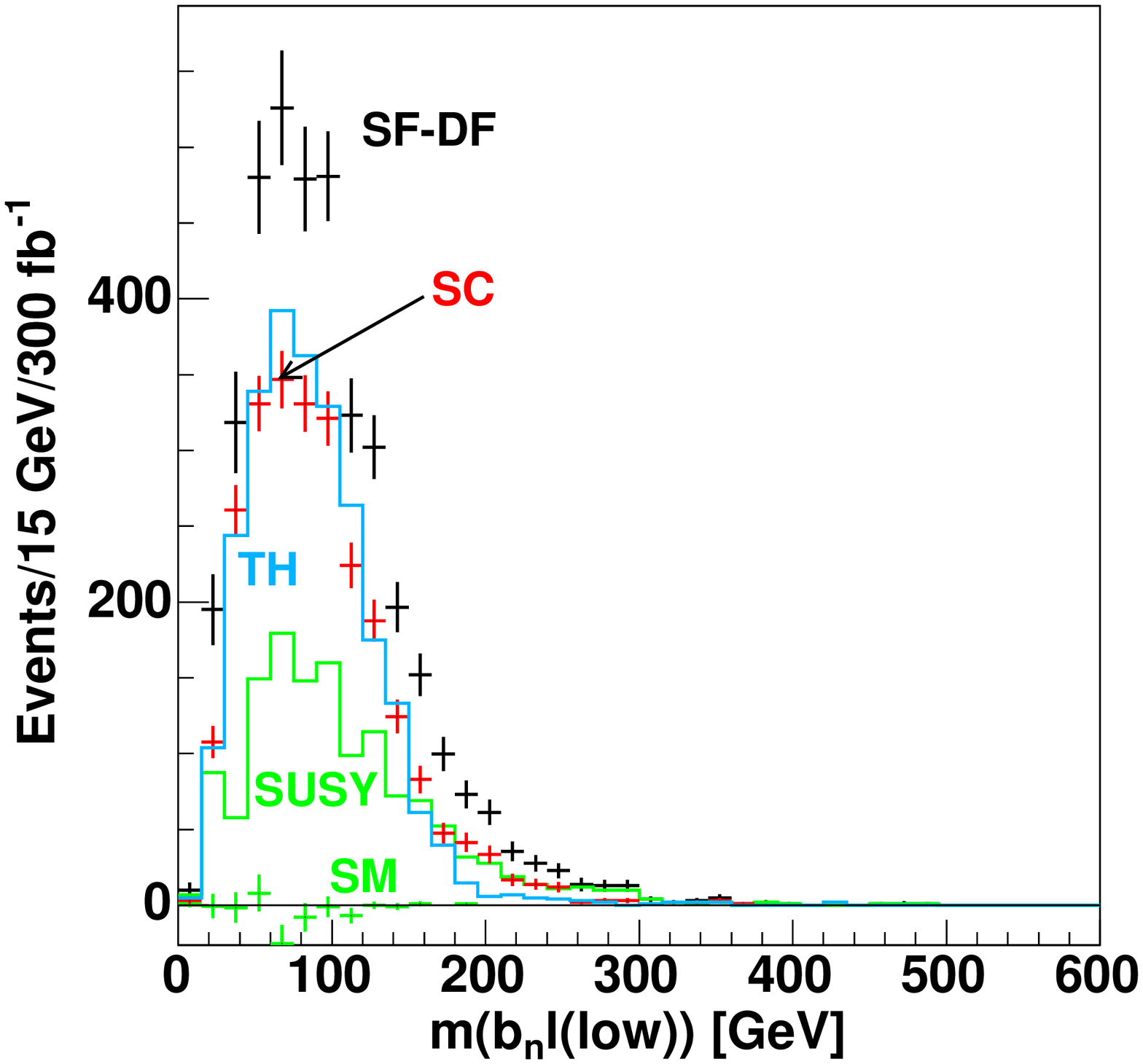}
} } }
\centerline{{\ifx\picnaturalsize N\epsfxsize \picsize\fi
\epsfbox{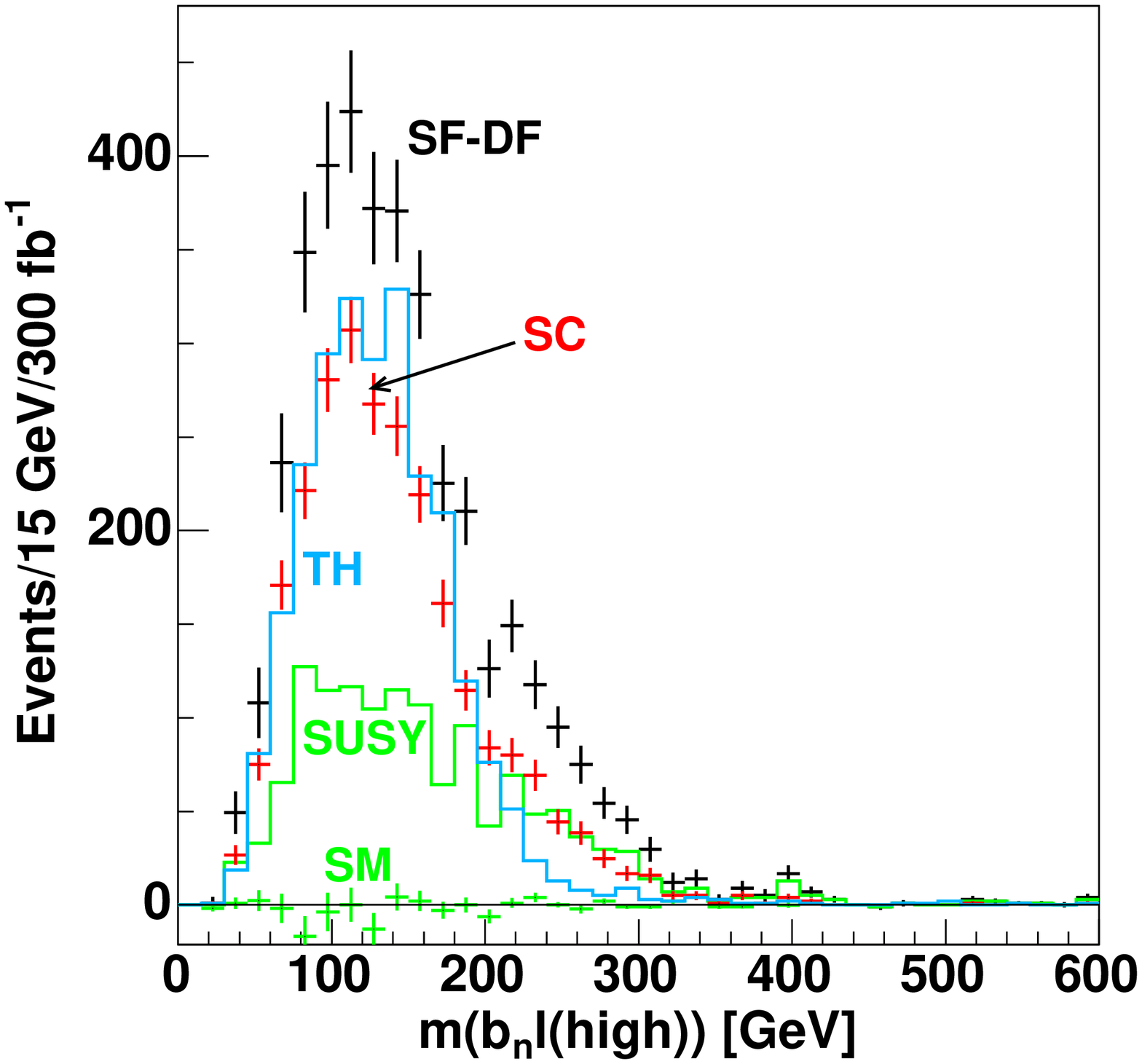}
} }
}\fi
\vspace*{-6mm}
\caption{Three bad $b$-tagged invariant mass distributions. See the 
text for details. \label{fig:mbb_2}}}

For the distributions in Fig.~\ref{fig:mbb_2} no endpoint estimation is 
attempted. 
The two $\mbNl$ distributions are distrusted since they
do not usually `point' towards the maximum value, 
as shown in Fig.~\ref{fig:TheoryDistr}. 
The phase space corresponding to the largest values is very small, 
leaving only an experimentally 
undetectable tail to mark the endpoint. 
These distributions are furthermore made less useful by 
the need to distinguish $\bN$ from $\bF$. For the distributions of 
Fig.~\ref{fig:mbb_1} this is not an issue because 
both $b$'s were used. 
At SPS~1a the endpoints involving $\bN$ have smaller nominal value than 
the corresponding endpoints with $\bF$.
Consequently, if we are not able to cleanly distinguish $\bN$ from $\bF$, 
all distributions which have $\bN$ as the only 
quark will get a contamination which stretches beyond the endpoints. 
If this contamination is substantial, and here it is, the $\bN$ endpoints will 
be washed out. 
The combination of these two effects seems to disable
these three distributions. 
The nominal endpoints are 
$\maxmbNll=281$~GeV, $\maxmbNlLow=197$~GeV and $\maxmbNlHigh=248$~GeV, 
which are clearly not obtainable from the distributions in 
Fig.~\ref{fig:mbb_2}.  
To cope with the problem of distinguishing $\bN$ and $\bF$ more general 
distributions could be constructed,
as was discussed in Sect.~\ref{subsect:quarkambiguity}.

For $\mbb$ the fit values are a considerable 25~GeV higher than the nominal 
value, and are not really accounted for by the errors. 
It is not completely understood why the discrepancy is so large; 
whether it is a statistical fluctuation or some other effect. 
Since the reconstruction resolution is much worse for jets than for leptons, 
one should expect that a mass distribution constructed from jets alone will be 
more smeared and therefore give worse endpoint determination than one which 
also involves leptons. This is probably part of the reason.

To extract the endpoints from the distribution, signal and background 
hypotheses are needed. 
Usually, at this level of detail, a straight line is used for the signal 
in the edge region. 
We see from the theory distribution of Fig.~\ref{fig:TheoryDistr} that 
the straight-line hypothesis in the edge region is normally quite well 
supported for the four distributions of Fig.~\ref{fig:mbb_1}. 
(The exception is $\mqqlHigh$, which in some cases might have a dangerous 
foot at the very last part of the edge.) 

Which background hypothesis to use is less clear. The background is here 
mainly from other SUSY processes, and therefore in principle unknown. 
One attempt to sketch the shape of the background, is by combining the lepton 
and the jet sectors of different events into a mixed-event sample, 
as described in Sect.~5 of \cite{Gjelsten:2004}. The shapes of the 
mixed-event distributions are then used as background hypotheses. 
Another way is to simply select for the background some appropriate function, 
based on the distribution slightly beyond the edge, where 
only background resides. 
An exponential or a polynomial usually gives a fair description. 
Whatever the signal and background hypotheses, systematics are introduced 
which may not always be easy to estimate.

For the plots of Fig.~\ref{fig:mbb_1} the signal was fitted with 
a straight line, while the background was modeled by both an
exponential and a mixed-event sample.
A four/three-parameter fit was then
performed by use of MINUIT \cite{James:1975dr} for exponential/mixed-event 
background
hypothesis and several different histogram binnings and fit ranges. 
Each distribution gives fit values and statistical
errors within a fairly narrow interval, indicative of
the systematic uncertainty in the fitting procedure.  The
results of the fits are summarised in Table~\ref{table:endpoints} at the end
of this section, after the discussion of energy scale errors.

\subsection{Propagation of energy scale errors\label{subsect:errorscale}}
In ATLAS it is expected that the absolute energy scale of jets and leptons 
will be known to 1\% and 0.1\%, respectively, see Ch.~12 of \cite{AtlasTDRvol1}.
This energy scale uncertainty translates into an uncertainty, 
or `error', for any mass constructed from jets and/or leptons. In particular,
the masses which go into our distributions have such an error in addition to 
the statistical error. 

Invariant masses which are constructed from jets alone or leptons alone, 
will inherit their uncertainties of 1\% or 0.1\%, respectively.  
For an invariant mass made from one lepton and one jet, the energy scale 
error is at 0.5\%. 
For masses constructed from more than two jets and leptons, the energy 
scale error is not constant, but varies within a given calculable interval, 
depending on whether it is the jet(s) or the lepton(s) which dominate 
the invariant mass for the given event. 

To find appropriate energy scale errors, we must investigate
$\mqqll$, $\mqql$ and $\mqll$. The overall masses can be expressed in
terms of two-particle masses, which have a constant energy scale error:
\begin{eqnarray}
\mqqll^2 &=& \mqq^2 + \mqNlN^2 + \mqNlF^2 + \mqFlN^2 + \mqFlF^2 + \mll^2 
\\
\mqql^2 &=& \mqq^2 + \mqNl^2 + \mqFl^2
\\
\mqll^2 &=& \mqlN^2 + \mqlF^2 + \mll^2
\end{eqnarray}
Absolute lower/upper limits for the energy scale error of the quantities on the 
left-hand side can then be found by assuming that the right-hand side is totally 
dominated by the term which has the smallest/largest energy scale error. 
This results in 
\begin{eqnarray}
\frac{\sigma(\mqqll)}{\mqqll} \in (0.1,1)\%
,\quad
\frac{\sigma(\mqql)}{\mqql} \in (0.5,1)\%
,\quad
\frac{\sigma(\mqll)}{\mqll} \in (0.1,0.5)\%
\label{eq:energyscaleranges}
\end{eqnarray}
where $\sigma$ denotes the energy scale error. 
These are absolute limits valid for any mass scenario. 
To find the relevant numbers for SPS~1a, all accepted events were reexamined;
scaling the jet momenta of each event by 1.01 and taking the ratio of the 
new and the old invariant mass. 
We then find the following average and root-mean-square values (in parentheses) 
of the relevant energy scale errors, 

\begin{eqnarray}
\frac{\sigma(\mbbll)}{\mbbll}       = 0.66 (0.10)\%,\quad
& &
\frac{\sigma(\mbblLow)}{\mbblLow}   = 0.80 (0.11)\%
\nonumber
\\
\frac{\sigma(\mbblHigh)}{\mbblHigh} = 0.71 (0.10)\%,\quad
& &
\frac{\sigma(\mbNll)}{\mbNll}       = 0.42 (0.02)\%
\label{eq:energyscaleerror}
\end{eqnarray}
Inclusion of the lepton energy scale will give a small correction to these numbers. 
For $\mbNll$, which we will not be using, the energy scale is nearly constant. 
For the three other distributions, the errors lie between approximately 0.5\% and 1\%, 
and fairly uniformly distributed, as is reflected in the root-mean-square values. 
For the fitting, only the energy scale error for masses 
which lie in the edge region is relevant. However, it turns out that at SPS~1a
and for these distributions, the error is almost the same for low as for 
high invariant mass values. 
Although the error of each distribution is found to lie in a fairly broad interval 
rather than being constant for all events, we have here used the average values,
Eq.~(\ref{eq:energyscaleerror}),
as a basis for the energy scale errors in Table~\ref{table:endpoints}.

An alternative approach could be to scale the jet momenta up/down as done above, 
then redo the entire fitting process on the new distributions and from this 
extract the effect of the energy scale error. 
For this to work one would have to disentangle the effect of the scaling from 
the yet uncontrolled systematics of the fitting procedure. 
At the present stage of fitting competence the gain from using a more correct 
procedure is probably lost in the increased complexity.

\begin{TABLE}{
\caption{Endpoint values found from fitting the edges in 
Fig.~\ref{fig:mbb_1}, for 300~$\text{fb}^{-1}$. 
The nominal values correspond to the mass of $\bO$ which  
is produced at significantly higher rates than the heavier $\bT$. 
The energy scale errors are based on the discussion in  
Sect.~\ref{subsect:errorscale}.
No values are given for the three distributions 
in Fig.~\ref{fig:mbb_2}.
}
\label{table:endpoints}
\begin{tabular}{|lcccc|}
\hline
  & Nominal & Fit & Energy Scale & Statistical \\
Edge & Value  & Value &Error ($\sigma^\scale$) & Error ($\sigma^\stat$) \\
  &  [GeV] & [GeV] & [GeV] & [GeV] \\
\hline
$\maxmbb$        & 312.7   & 335--339   &  3.4  & 6--10  \\
$\maxmbbll$      & 496.3   & 494--500   &  3.3  &  5--7  \\
$\maxmbblLow$    & 413.2   & 407--417   &  3.3  & 8--12  \\
$\maxmbblHigh$   & 461.9   & 454--462   &  3.3  &  5--7  \\
\hline
\end{tabular} }
\end{TABLE}

\section{Masses from endpoints}\label{sect:masses}

\subsection{10,000 ATLAS experiments}

The sparticle masses can be obtained from a numerical fit based 
on the gluino endpoints found in the previous section together with 
the squark endpoints of \cite{Gjelsten:2004}. 
Due to the simplistic treatment of the detector effects and the 
endpoint fitting, we do not trust the fit values to be representative 
of what a more complete simulation would give. 
Therefore, rather than use directly the endpoint values obtained from 
the distributions, we generate an ensemble of 10,000 ATLAS experiments 
based on the estimated {\it errors}, as done in \cite{Allanach:2000kt}. 

For each experiment the endpoint values $\vvec{E}^\exp$ 
are combined with the general endpoint expressions 
$\vvec{E}^\theory$ 
in a least-square function $\LSfun$, 
\begin{equation}
\LSfun = 
[\vvec{E}^\exp-\vvec{E}^\theory(\vvec{m})]^T 
\mat{W}
[\vvec{E}^\exp-\vvec{E}^\theory(\vvec{m})]
\label{eq:LSfun}
\end{equation}
The weight matrix $\mat{W}$ is the inverse of the covariance matrix 
which is constructed from the endpoint errors and appropriately 
handles the endpoint correlations due to the energy scale error. 
The minimisation of $\LSfun$ then yields the masses. 
Due to the composite nature of the endpoint expressions, there are
usually several competing $\LSfun$ minima for a given set of endpoints. 
If these minima are close in $\LSfun$ value, they must all be considered, 
opening the way for multiple mass solutions. 
Finally, ensemble distributions for the masses can be plotted and studied, 
and are to be interpreted as probability distributions for the 
masses which can be obtained at the LHC.  
We will be interested in the mean and width of these ensemble distributions. 
For more details of the procedure, see \cite{Gjelsten:2004}.

\subsection{Mass estimation}
In total six sparticles are involved in our decay chains: 
$\NO$, $\lR$, $\NT$, $\qL$, $\bO$ and $\gl$.
The masses of the first five of these could already be obtained 
from the squark endpoints alone, as was done in \cite{Gjelsten:2004}. 
The gluino mass is accessed by 
the gluino endpoints. 
%
Since these also involve the other masses (except $\mqL$), 
the squark endpoints and the gluino endpoints 
are combined in a single fit where all the masses are free parameters. 
The six squark-endpoint measurements are given in Table~4 
(upper part) of \cite{Gjelsten:2004}.
The values of the right-most column, `Syst.~Fit Error' are not used. 
For the gluino endpoints of Table~\ref{table:endpoints} (in the present paper)
the statistical 
errors are given as intervals. We take here the midpoints of the intervals,  
and further assume that the systematics of the fit values (also given 
by intervals) will be settled, or at least dominated by the statistical 
errors. This assumption might be somewhat optimistic, and should be 
kept in mind when contemplating the results obtained below. 
(In the next section we will look at the effects of doubling the errors.)
Furthermore, we decided to use only three out of seven gluino 
distributions. In particular the choice of excluding $\maxmbb$ 
was made after 
consultation with the Monte Carlo truth at our specific SUSY scenario, 
which is not a viable strategy in a realistic setting. 
More study could however promote this result into a generic one.

If we start from the squark endpoints and add only {\it one} of the 
gluino endpoints in the numerical fit, the gluino mass will be 
determined and all the other masses will remain unchanged. 
This is so because there is exactly one new measurement for one new mass. 
The gluino mass returned from the fit is the one which gives zero 
for the new terms added to $\LSfun$, so there is no increase in the 
$\LSfun$-value. 
Also, the number of minima remains unchanged by adding one more 
measurement for one more mass.
This situation is similar to the determination of $\mbO$ in \cite{Gjelsten:2004}. 
Only one endpoint involves $\bO$, so its inclusion has no effect on 
the other masses or number of minima. 

When more gluino endpoints are added, the gluino sector 
becomes over-determined, and the positions of the minima will change, 
i.e.\ the other masses will be affected. Since the gluino endpoints  
have somewhat larger errors than the squark endpoints, large effects 
are not expected, except perhaps for $\bO$. 
Below, results are given for the case when all three selected gluino 
endpoints are used.

\begin{TABLE} {
\caption{Number of minima for various $\Delta\LSfun$ cuts and 
their whereabouts.\label{table:freq}} 
\begin{tabular}{|l|c|cc|}
\hline
  & $\#$ Minima & {\itB(1,1)} & {\itB(1,2)}  \\
\hline 
$\Delta\LSfun= 0$ & 1.00 & 91\%  &  9\% \\
$\Delta\LSfun\leq 1$ & 1.11 & 95\%  & 16\% \\
$\Delta\LSfun\leq 3$ & 1.30 & 98\%  & 32\% \\
$\Delta\LSfun\leq99$ & 1.87 & 99\%  & 87\% \\
\hline
\end{tabular}
}
\end{TABLE}

For the current precision of the endpoint measurements, the 
numerical fit nearly always returns two minima, one in mass 
region {\itB (1,1)},\footnote{The definition of the regions 
is given in Sect.~4.3 of \cite{Gjelsten:2004}. For a mass region 
{\itB{(i,j)}} the first index refers to the expression used for 
$\maxmqFll$, the second index shows which combination is used for 
($\maxmqFlLow$,$\maxmqFlHigh$), see Eqs.~(4.4)--(4.5) of 
\cite{Gjelsten:2004}.}
which is the region of the nominal 
masses at SPS~1a, and one in mass region {\itB(1,2)}. 
If the minima are close in $\LSfun$-value, both must be considered. 
Table~\ref{table:freq} shows the probability of having 
more than one solution in an experiment. 
The cut on $\Delta\LSfun$, the distance to the global minimum, 
gives the quality of the second minimum.
In most cases the {\itB(1,1)} minimum is the selected one. 
These numbers are very similar to the numbers obtained without the gluino
endpoints, Table~5 of~\cite{Gjelsten:2004}, where a more detailed
description can also be found.

\begin{TABLE} {
\def\spcA{\mbox{\hspace{0ex}}}
\begin{tabular}{|c|r|rr|rr|}
\hline
  &   &\multicolumn{2}{|c|}{\itB(1,1)}  &\multicolumn{2}{|c|}{\itB(1,2)} \\
  & Nom & Mean & RMS & Mean & RMS \\
\hline 
$\mNO     $ &   96.1  &   96.3 &  3.7 &   85.5 &  3.4  \\
$\mlR     $ &  143.0  &  143.2 &  3.7 &  130.6 &  3.8  \\
$\mNT     $ &  176.8  &  177.0 &  3.6 &  165.7 &  3.5  \\
$\mqL     $ &  537.2  &  537.5 &  6.0 &  523.5 &  5.0  \\
$\mbO     $ &  491.9  &  492.2 & 12.5 &  471.8 & 12.6  \\
$\mgl     $ &  595.2  &  595.5 &  7.2 &  582.5 &  6.8  \\
\hline
$\mlR-\mNO$ &   46.92  &   46.93 &  0.28 &   45.11 &  0.72 \\ 
$\mNT-\mNO$ &   80.77  &   80.77 &  0.18 &   80.19 &  0.29 \\ 
$\mqL-\mNO$ &  441.2\spcA  &  441.2\spcA &  3.1\spcA &  438.0\spcA &  2.8\spcA  \\
$\mbO-\mNO$ &  395.9\spcA  &  396.0\spcA & 11.2\spcA &  386.3\spcA & 11.2\spcA  \\
$\mgl-\mNO$ &  499.1\spcA  &  499.2\spcA &  5.6\spcA &  497.0\spcA &  5.4\spcA  \\
$\mgl-\mbO$ &  103.3\spcA  &  103.3\spcA &  9.1\spcA &  110.7\spcA &  9.5\spcA  \\
\hline
\end{tabular}
\caption{Masses (Mean) and root-mean-square deviations from the mean
(RMS) of minima in regions {\itB(1,1)} and {\itB(1,2)}, for 
$\Delta\LSfun\leq1$. The nominal masses (Nom) are given in the second column. 
All values in GeV. See the text for more details. 
\label{table:masses}} 
}
\end{TABLE}

The masses are given in Table~\ref{table:masses} for minima which 
satisfy $\Delta\LSfun\leq1$. 
The numbers are very close to the ones obtained without 
the gluino endpoints, see Table~6 of~\cite{Gjelsten:2004}. 
Only $\bO$ is affected, as was expected. 

The gluino mass is quite well determined. The ensemble mean is at  
the nominal value, and the root-mean-square deviation from the mean
is 7.2~GeV, only a GeV more than for $\mqL$. 
For the other masses, especially the lighter ones, 
mass differences are much more accurately determined than the masses 
themselves. 
This is also the case for the gluino, although to a lesser degree,
as is seen from the smaller root-mean-square value for $\mgl-\mNO$ 
of 5.6/5.4~GeV. 
When the correlation to $\NO$ is taken into account, as in mass 
differences, $\qL$ is better determined than $\gl$, as it naively 
should be from the smaller squark endpoint errors.  
Since the gluino endpoints involve the sbottom mass, one might 
expect that the masses of $\gl$ and $\bO$ have a considerable correlation. 
Some correlation is found, as can be seen from comparing the root-mean-square 
value of $\mgl-\mbO$ with that of $\mbO$ alone.

\FIGURE[ht]{
\epsfxsize 16.3cm
\centerline{\epsfbox{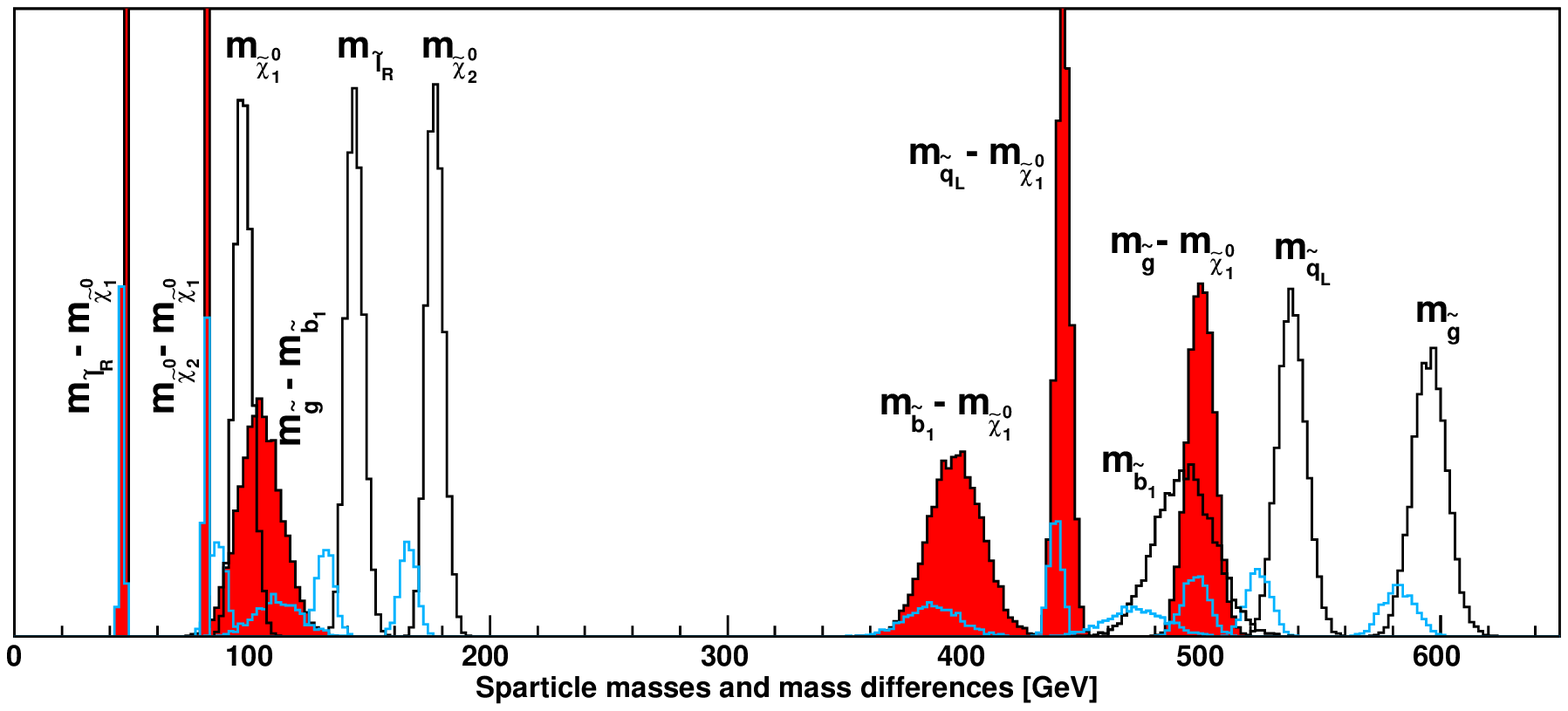}}
\vspace{-0.6cm}
\caption{Sparticle masses and mass differences for solutions with
$\Delta\Sigma\leq1$. See the text for details. 
\label{fig:masses}}}

The ensemble distributions summarised in Table~\ref{table:masses}, 
i.e.\ for $\Delta\LSfun\leq1$, are shown in Fig.~\ref{fig:masses}. 
The unfilled distributions in black show from left to
right $\mNO$, $\mlR$, $\mNT$, $\mbO$, $\mqL$ and $\mgl$ 
for solutions in the nominal region {\itB(1,1)}. 
We will have such a solution in 95\% of the experiments, 
see Table~\ref{table:freq}.
The unfilled distributions in blue (grey) show the same masses
for solutions in region {\itB(1,2)}. 
Such a solution occurs in 16\% of the experiments.  
(For $\Delta\LSfun\leq1$ there is an 11\% chance of having both solution 
types.)
The smaller rate of the {\itB(1,2)} solutions is reflected in the smaller 
area under the blue curves.  
The {\itB(1,2)} solutions return slightly lower masses, so these distributions 
are positioned `down' (due to the lower rate) and slightly to the left 
of the {\itB(1,1)} distributions. 
The filled distributions show from left to right
$\mlR\!-\!\mNO$, $\mNT\!-\!\mNO$, 
$\mgl\!-\!\mbO$,
$\mbO\!-\!\mNO$, $\mqL\!-\!\mNO$ and
$\mgl\!-\!\mNO$.  
Again, the most populated distributions (black curves) are for solutions 
in region {\itB(1,1)}, the least populated (blue curves) for {\itB(1,2)} 
solutions.
For mass {\it differences} there is more overlap between the {\itB(1,1)} and 
{\itB(1,2)} solutions, in particular for $\mlR\!-\!\mNO$ and $\mNT\!-\!\mNO$, 
of which only the lower parts of the distributions are visible. 
Mass differences are better determined than the masses themselves, 
reflected here by the narrower distributions of the former. 
In the case of $\bO$ this effect is clear for $\mgl-\mbO$,
much less so for $\mbO-\mNO$, which points to the fact that
$\bO$ largely decouples from $\NO$,

\subsection{Edge sensitivities}
It may seem a puzzle that $\mgl$ is determined 
with much higher precision than $\mbO$. 
Both masses enter the three new gluino endpoints, but 
$\mbO$ has in addition the $\minmbFllThres$ measurement. 
Naively one might therefore expect the $\bO$ mass to be more precisely determined. 

In order to better see how masses are constrained by the endpoint formulas, 
consider the part of the least-square function $\LSfun$ which involves one 
given endpoint and one given mass only. 
For simplicity of argument, assume that the endpoint 
measurement is at the nominal value, as are the masses. 
Changing the given mass away from the nominal value by an amount $\Delta m$
will give an increase of our selected $\LSfun$ part by 
\begin{eqnarray}
\Big(\frac{E(m+\Delta m)-E^\exp}{\sigma_{E}}\Big)^2 
\approx \Big(\frac{E(m)+(\partial E/\partial m)\Delta m-E^\exp}{\sigma_{E}}\Big)^2 
=\Big(\frac{(\partial E/\partial m)\Delta m}{\sigma_{E}}\Big)^2 
\end{eqnarray}
where $E$ is the value of a particular kinematical endpoint.
The value of $\Delta m$ which gives an increase of $\LSfun$ by 1 (from this term only), 
is then given by
\begin{equation}
\Delta m=\frac{\sigma_{E}}{(\partial E/\partial m)}
\end{equation}

\def\Bpart{\frac{\partial E}{\partial m}}\def\Bmass{\Delta m}
\begin{TABLE} {
\begin{tabular}{|c|rr|rr|rr|rr|rr|rr|}
\hline
&\multicolumn{2}{|c|}{$\NO$} &\multicolumn{2}{|c|}{$\lR$} &\multicolumn{2}{|c|}{$\NT$} 
&\multicolumn{2}{|c|}{$\qL$} &\multicolumn{2}{|c|}{$\bO$} &\multicolumn{2}{|c|}{$\gl$} \\
& $\Bpart$ & $\Bmass$ & $\Bpart$ & $\Bmass$ & $\Bpart$ & $\Bmass$ & $\Bpart$ & $\Bmass$ & $\Bpart$ & $\Bmass$ & $\Bpart$ & $\Bmass$  \\
\hline
$\maxmll$  &   -0.7 &   -0.1  &   -0.6 &   -0.2  &    1.3 &    0.1  &    -   &    -    &    -   &    -    &    -   &    -   \\
 
$\maxmqFll$  &   -1.9 &   -1.2  &    -   &    -    &    0.7 &    3.2  &    0.9 &    2.6  &    -   &    -    &    -   &    -   \\
$\minmqFllThres$  &   -0.8 &   -3.2  &   -1.6 &   -1.5  &    1.7 &    1.4  &    0.4 &    6.2  &    -   &    -    &    -   &    -   \\
$\maxmqFlLow$  &    -   &    -    &   -3.9 &   -0.4  &    3.0 &    0.6  &    0.6 &    2.8  &    -   &    -    &    -   &    -   \\
$\maxmqFlHigh$  &   -3.2 &   -0.7  &    2.2 &    1.0  &   -0.3 &   -8.2  &    0.8 &    2.7  &    -   &    -    &    -   &    -   \\
$\minmbFllThres$  &   -0.7 &   -6.5  &   -1.5 &   -3.1  &    1.5 &    3.0  &    -   &    -    &    0.4 &   11.8  &    -   &    -   \\
 
\hline
$\maxmbb$  &    -   &    -    &    -   &    -    &   -0.3 &  -39.6  &    -   &    -    &   -1.3 &   -8.1  &    1.7 &    6.3 \\
$\maxmbbll$  &   -1.3 &   -5.2  &    -   &    -    &    0.3 &   23.4  &    -   &    -    &   -0.1 &  -65.1  &    1.0 &    6.5 \\
$\maxmbblLow$  &    -   &    -    &   -2.3 &   -4.5  &    1.5 &    6.8  &    -   &    -    &   -0.6 &  -19.0  &    1.3 &    8.4 \\
$\maxmbblHigh$  &   -2.1 &   -3.2  &    1.4 &    4.7  &   -0.4 &  -17.6  &    -   &    -    &   -0.3 &  -24.5  &    1.1 &    6.1 \\
 
\hline
$\maxmbNll$  &   -1.2 &    -    &    -   &    -    &    0.7 &    -    &    -   &    -    &   -1.2 &    -    &    1.5 &    -   \\
$\maxmbNlLow$  &    -   &    -    &   -2.6 &    -    &    2.1 &    -    &    -   &    -    &   -0.9 &    -    &    1.0 &    -   \\
$\maxmbNlHigh$  &   -2.1 &    -    &    1.4 &    -    &    -   &    -    &    -   &    -    &   -1.1 &    -    &    1.3 &    -   \\
\hline
\end{tabular}
\caption{Partial derivatives of endpoints with respect to masses 
at the nominal mass values. 
The larger the partial derivative $\partial E/\partial m$ is, 
the more sensitive the given endpoint is to the given mass. 
The quantity $\Delta m$ [GeV] is defined by $\sigma_{E}/(\partial E/\partial m)$, 
where $\sigma_{E}$ is the combined statistical and energy scale error of the endpoint.  
Numbers are specific for SPS~1a. 
See the text for more details.
\label{table:derivatives}} }
\end{TABLE}

Since the shift in mass also induces changes to other contributions to
$\LSfun$, the interpretation of $\Delta m$ is not straight-forward in terms of
mass error.  Still, $\Delta m$ does combine the precision of the endpoint,
given by the endpoint error $\sigma_{E}$, with the response of the endpoint to
the mass, which are the two important quantities.  The same relation could be
derived in a mathematically more intuitive way,
\begin{equation}
\Delta m = \Big(\frac{\partial m}{\partial E}\Big) \Delta E =
\frac{\sigma_{E}}{(\partial E/\partial m)}
\end{equation}
where the endpoint variation $\Delta E$ is set equal to the experimental error
$\sigma_{E}$. 

Inclusion of the endpoint error opens for comparison between different
endpoints, even if only in a semi-quantitative way which does not take
correlations into account. The importance of the different endpoints in
constraining a given mass can then be studied. In
Table~\ref{table:derivatives}, values for $\partial E/\partial m$ and $\Delta
m$ are given for both the squark and the gluino endpoints. For the three
gluino endpoints of Fig~\ref{fig:mbb_2}, where no fit was made and hence no
$\sigma$ is available, $\Delta m$ is not given.

These numbers show immediately that the three gluino endpoints included in the
numerical fit have a strong sensitivity to the gluino mass, but only a very
moderate sensitivity to the sbottom mass. Also $\minmbFllThres$ is seen not to
constrain $\mbO$ too much.  If $\maxmbb$ could be used or some of the squark
endpoints other than $\minmbFllThres$ (numbers shown only
for~$\minmqFllThres$), then $\mbO$ might be determined more accurately.
The table can also be used to understand better which of the endpoints are
important in constraining each of the lighter masses.

\section{Increasing the errors}\label{sect:doublederror}
In the previous section the uncertainties of sparticle masses were found from
the estimated errors of the endpoint measurements.  The technique of
generating an ensemble of experiments with endpoint measurements spread
Gaussian-like around the nominal values sidesteps some of the problems of the
crude detector simulation and edge-fitting procedure, as only the estimated
statistical errors and energy scale errors are then relevant.  However, these
are also subject to uncertainties.  First, for the statistical errors we used
the means of supposedly reasonable intervals obtained from recurrent fits with
slight variations in bin width, range of fit and fit functions. Clearly these
numbers are tentative.  Second, one cannot be certain that the energy scale
will actually be known to the precision aimed for.

In this section we will therefore investigate two somewhat more conservative
scenarios.  First we double the statistical errors of all the endpoints,
keeping the energy scale errors at the `default', i.e.\ at the values given in
Table~\ref{table:endpoints} (this paper) and Table~4 of \cite{Gjelsten:2004}.  
Then we keep the statistical errors at the
default values and double instead the energy scale errors.

However, in order to better understand the changes this will lead to, 
we first try to disentangle the impact of the statistical errors 
from that of the energy scale errors for a simplified version of 
the default scenario considered in the previous section.

\subsection{Disentangling statistical from energy scale errors}
\label{sect:disentangle}

A simplified method is considered, in which only the four endpoints 
$\maxmll$, $\maxmqFll$, $\maxmqFlLow$ and $\maxmqFlHigh$ are  used 
to obtain the masses of $\NO$, $\lR$, $\NT$ and $\qL$. 
The main features of the comprehensive method where all the endpoints are used, 
are also present for this simplified analysis.
In particular we can use the inversion formulas of \cite{Gjelsten:2004}. 
For region {\itB(1,1)} Eqs.~(4.13)--(4.16) are appropriate. 
Notice in these formulas how the three lightest sparticle masses are 
proportional to $\maxmll$.
Hence, these three masses scale with the lepton energy scale error. 
Conversely, compare this with their dependence on the three squark endpoints. 
If these three endpoints have identical relative energy scale errors 
[which in reality is only approximately true for $\maxmqFll$ -- here we have 
assumed it to be exactly true], 
then the energy scale errors cancel out for the three lightest masses, 
as is easily seen by letting each of the squark endpoints be scaled by a 
factor $(1+\delta)$. 
Consequently, the impact of the energy scale errors on the three lightest 
sparticles is equal to the lepton energy scale error, 
which for most purposes is negligible. 
For the squark mass, this symmetry is broken, 
$\mqL$ is not proportional to $\maxmll$, 
so the other relations do not apply either. The squark mass therefore has a 
noticeable dependence on the energy scale errors. 

\begin{TABLE}[htb] {
\def\ph{\phantom}
\begin{tabular}{|c|c|c|c|}
\hline
 & Only  & Only  & Both \\ & $\sigma^\stat$ & $\sigma^\scale$ &  \\
\hline
$\mNO$ & 3.9\ph{0} & 0.10 & 3.9\ph{0} \\
$\mlR$ & 3.9\ph{0} & 0.14 & 3.9\ph{0} \\
$\mNT$ & 3.8\ph{0} & 0.18 & 3.8\ph{0} \\
$\mqL$ & 5.9\ph{0} & 2.4\ph{0} & 6.3\ph{0} \\
\hline
$\mlR-\mNO$ & 0.28& 0.05 & 0.28 \\
$\mNT-\mNO$ & 0.16& 0.08 & 0.18 \\
$\mqL-\mNO$ &  2.0\ph{0} & 2.4\ph{0} & 3.1\ph{0} \\
\hline
\end{tabular}
\caption{Disentangling statistical from energy scale errors. 
Root-mean-square values of uncertainties in masses 
and mass differences [GeV]. 
Only the four basic endpoints have been used.\label{table:masses-simplistic}} }
\end{TABLE}

In order to quantify the statements just made, an ensemble of ATLAS
experiments was constructed in the usual way \cite{Gjelsten:2004}, using the
default errors of the four squark endpoints.
Table~\ref{table:masses-simplistic} shows the root-mean-square values thus
obtained for the four masses and the mass differences.  Three cases are shown:
only statistical errors, only energy scale errors and combined.
Even though the energy scale errors are roughly twice the statistical errors
for the endpoints being used (Table~4 of \cite{Gjelsten:2004}), the latter
clearly dominate the mass errors.  Only for combinations involving $\mqL$ do
the energy scale errors play a role.  For $\mqL-\mNO$ the impact of the two
uncertainties are comparable.  Comparison with Table~\ref{table:masses} shows
that the stripped scenario accounts for the main features of the complete
scenario to which we return below.

\subsection{Doubling the statistical errors}

\begin{TABLE} {
\begin{tabular}{|l|c|cc|}
\hline
  & $\#$ Minima & {\itB(1,1)} & {\itB(1,2)}  \\
\hline
$\Delta\LSfun= 0$ & 1.00 & 73\%  & 27\% \\
$\Delta\LSfun\leq 1$ & 1.24 & 84\%  & 41\% \\
$\Delta\LSfun\leq 3$ & 1.47 & 90\%  & 57\% \\
$\Delta\LSfun\leq99$ & 1.87 & 93\%  & 75\% \\
\hline
\end{tabular}
\caption{Number of minima for various $\Delta\LSfun$ cuts and 
their whereabouts. The statistical errors are doubled,
compared to Table~\ref{table:freq}.
\label{table:freq2stat}}  }
\end{TABLE}

Having established that the statistical errors dominate the errors on the
masses, we expect large effects from doubling these errors.
Table~\ref{table:freq2stat} shows a noticeable increase in the rate of
{\itB(1,2)} solutions relative to the nominal {\itB(1,1)} solutions.  This is
no surprise, as SPS~1a lies very close to the border.  Allowing larger errors
for the endpoints naturally allows more of the {\itB(1,2)} region to be
accessed.  Furthermore, there is an increase in the number of solutions, which
is also reasonable.  The 10--15~GeV lower mass values of the {\itB(1,2)}
solutions become more troubling as the occurrence of this minimum becomes more
frequent.

\begin{TABLE} {
\begin{tabular}{|c|r|rc|rc|}
\hline
  &  & \multicolumn{2}{|c|}{\itB(1,1)} &\multicolumn{2}{|c|}{\itB(1,2)} \\
  & Nom & Mean & RMS & Mean & RMS  \\
\hline
$\mNO    $ &   96.1  &   97.5 &  7.2 &   84.2 &  6.1  \\
$\mlR    $ &  143.0  &  144.5 &  7.0 &  129.0 &  6.7  \\
$\mNT    $ &  176.8  &  178.3 &  7.0 &  164.3 &  6.2  \\
$\mqL    $ &  537.2  &  539.3 & 11.0 &  522.1 &  8.7  \\
$\mbO    $ &  491.9  &  494.4 & 24.3 &  468.6 & 24.0  \\
$\mgl    $ &  595.2  &  597.4 & 12.9 &  581.1 & 11.7  \\
\hline
$\mlR-\mNO$ &  46.92  &  47.03 & 0.53  &  44.72 & 1.21  \\
$\mNT-\mNO$ &  80.77  &  80.80 & 0.32  &  80.07 & 0.48  \\
$\mqL-\mNO$ &  441.2  &  441.8 &  4.4  &  437.8 &  3.7  \\
$\mbO-\mNO$ &  395.9  &  396.9 & 21.9  &  384.4 & 21.9  \\
$\mgl-\mNO$ &  499.1  &  499.9 &  9.4  &  496.9 &  9.0  \\
$\mgl-\mbO$ &  103.3  &  103.0 & 18.0  &  112.5 & 18.9  \\
\hline
\end{tabular}
\caption{Same as Table~\ref{table:masses}, except for doubled {\it
statistical} errors: Mean and root-mean-square deviations from the mean (RMS)
[GeV].
\label{table:masses2stat}} }
\end{TABLE}

From Table~\ref{table:masses2stat} we see that doubling the statistical errors
very nearly amounts to doubling the errors on the masses.  The exception is
mainly $\mqL-\mNO$ whose error increases less.  This effect is explained by
the discussion in Sect.~\ref{sect:disentangle}, see
Table~\ref{table:masses-simplistic}, which shows that for this particular mass
combination the statistical errors only account for half the total error on
the mass difference.
For $\gl$ a pattern similar to, but less pronounced than that of $\qL$ can be
seen.

\subsection{Doubling the energy scale errors}

As already anticipated by the previous discussions, changes in 
the energy scale errors have only minor effects in the SPS~1a scenario 
at the given integrated luminosity. 
The doubling of the energy scale errors does not at all change the number 
and positions of minima from the default values of Table~\ref{table:freq}. 

\begin{TABLE} {
\label{table:}
\begin{tabular}{|c|r|rc|rc|}
\hline
  &  & \multicolumn{2}{|c|}{\itB(1,1)} &\multicolumn{2}{|c|}{\itB(1,2)} \\
  & Nom & Mean & RMS  & Mean & RMS  \\
\hline
$\mNO    $ &   96.1  &   96.3 &  3.7 &   85.5 &  3.4  \\
$\mlR    $ &  143.0  &  143.2 &  3.7 &  130.6 &  3.8  \\
$\mNT    $ &  176.8  &  177.0 &  3.6 &  165.7 &  3.5  \\
$\mqL    $ &  537.2  &  537.3 &  7.3 &  523.8 &  6.4  \\
$\mbO    $ &  491.9  &  492.1 & 13.0 &  472.1 & 13.1  \\
$\mgl    $ &  595.2  &  595.3 &  9.3 &  582.9 &  8.9  \\
\hline
$\mlR-\mNO$ &  46.92  &  46.93 & 0.29 &  45.11 & 0.72  \\
$\mNT-\mNO$ &  80.77  &  80.77 & 0.22 &  80.19 & 0.32  \\
$\mqL-\mNO$ &  441.2  &  441.1 &  5.1 &  438.3 &  4.9  \\
$\mbO-\mNO$ &  395.9  &  395.8 & 11.8 &  386.6 & 11.8  \\
$\mgl-\mNO$ &  499.1  &  499.0 &  8.1 &  497.4 &  8.1  \\
$\mgl-\mbO$ &  103.3  &  103.2 &  9.3 &  110.9 &  9.8  \\
\hline
\end{tabular}
\caption{Same as Table~\ref{table:masses}, except for doubled {\it energy
scale} errors: Mean and root-mean-square deviations from the mean (RMS) [GeV].
\label{table:masses2scale}} }
\end{TABLE}

Furthermore, as is seen from Table~\ref{table:masses2scale}, most masses
remain largely unaffected, the exception being combinations involving $\mqL$
and to a lesser extent $\mgl$, as previously noted. The explanation is the
same: For the lighter masses the jet energy scale errors cancel out, leaving
only the minimal effect of the lepton scale error.  In the case of $\mqL$ it
does play a role, although not dominant, giving a slight increase, and for
$\mqL-\mNO$ the impact of the energy scale errors is again significant.  The
same applies to $\mgl$ and $\mgl-\mNO$.

\section{LHC + Linear Collider (LC)\label{sect:LC}}
Within the time-frame of the analysis of LHC data, measurements from a 
Linear Collider may become available. 
While the LHC is able to measure mass differences at high precision, 
as documented in the previous sections, 
a Linear Collider will, due to the much cleaner environment, 
provide precise mass measurements of the sparticles which are kinematically 
accessible, in particular $\NO$. 
Such a measurement will be the scale fixer which is lacking in the 
LHC data, and will in combination with the LHC measurements allow one to also
fix the masses of the heavier sparticles not accessible at the Linear Collider. 

To estimate the effect of a Linear Collider measurement of $\mNO$, 
we add to our least-square function $\LSfun$, a term 
$[(\mNO-m_{\NO}^{\rm LC})/\sigma^{\rm LC}_{\mNO}]^2$, where
the quantity with superscript `LC' is the Linear Collider measurement, 
and generate an ensemble of LHC and LC experiments.  
Since $\sigma^{\rm LC}_{\mNO}=0.05$~GeV \cite{Aguilar-Saavedra:2001rg}, 
the LC measurement practically fixes $\mNO$ at the nominal value. 
(For the LHC measurements we use in this section the default endpoint errors.)

\begin{TABLE} { 
\def\spcA{\mbox{\hspace{1.2ex}}}
\label{table:massesLC}
\begin{tabular}{|c|r|rr|}
\hline
  &  &\multicolumn{2}{|c|}{\itB(1,1)} \\ 
  & Nom & Mean & RMS  \\
\hline
$\mNO    $ &   96.05  &   96.05 &  0.05 \\
$\mlR    $ &  142.97  &  142.97 &  0.29 \\
$\mNT    $ &  176.82  &  176.82 &  0.17 \\
$\mqL    $ &  537.2\spcA  &  537.2\spcA  &  2.5\spcA  \\
$\mbO    $ &  491.9\spcA  &  491.9\spcA  & 10.9\spcA  \\
$\mgl    $ &  595.2\spcA  &  595.2\spcA  &  5.5\spcA  \\
$\mgl-\mbO$&  103.3\spcA  &  103.3\spcA  &  9.0\spcA \\ 
\hline
\end{tabular}
\caption{Mass values (all in GeV) from LHC+LC. Since the occurrences 
of {\itB(1,2)} solutions are reduced to $\sim1$\%, they are left out.}}
\end{TABLE}

For the {\itB(1,2)} solutions, which normally return $\NO$ masses some 10~GeV
below the nominal value, the LC measurement has the dramatic effect of
reducing their occurrences to~$\sim1$\% (for $\Delta\LSfun\leq\nolinebreak3$).
These minima can therefore for most purposes be neglected.  As a consequence,
the probability of having two minima is strongly reduced, to the per mille
level.  For SPS~1a the Linear Collider measurement thus closes the issue of
multiple minima altogether.  (This behaviour was already reported in
\cite{Gjelsten:2004}, the inclusion of the gluino endpoints thus makes no
difference in this respect.)

The combined LHC + LC results are shown in Table~\ref{table:massesLC}.
Comparison with the numbers of Table~\ref{table:masses} shows
that the precision of the mass determination improves considerably 
when the LC measurement is included.
The root-mean-square values of the masses are now approximately equal to the 
root-mean-square values of mass differences without the LC measurement. 
For $\mbO$ the correlation to $\mNO$ is not dominant. 
The spread of $\mbO$ is therefore still larger than the 
spread of $\mgl-\mbO$, which does not feel the fixing of $\mNO$.
As was the case without the LC measurement, the inclusion of the 
gluino endpoints in the numerical fit does not affect the other masses. 
The exception is $\mbO$, for which the root-mean-square value is reduced 
by $\sim$1~GeV compared to the results in \cite{Gjelsten:2004}.

\section{Conclusion}\label{sect:conclusion}

In this paper we have extended the endpoint method of obtaining 
masses in R-parity conserving SUSY scenarios to also include the 
gluino mass, given the decay chain (\ref{eq:gluinochain}). 
All the gluino endpoints have been calculated. 
Details of the calculations of $\maxmqqll$ and $\maxmqFll$ are given 
in the appendices. 
Theory distributions for the seven new distributions were studied 
for a selection of mass scenarios, revealing that some of the 
distributions often have little phase space towards higher masses, 
making them less useful.

An ATLAS simulation of 300 fb$^{-1}$ was performed for the mSUGRA 
point SPS~1a. 
While we were not able to detect the gluino edges in the case of an 
intermediate first or second-generation squark ($\qL$), edges were 
found and fitted in the case of an intermediate sbottom. 
This is due to fact that 80\% of the $\sb$'s come from a gluino, 
together with the jet selection requirement of having two 
and only two $b$-tagged jets, 
both important background-reducing factors.
Not all of the endpoints were used. 
Energy scale errors were discussed and found to lie in intervals.  
Yet, they were taken as constants 
in the analysis.

To estimate the precision with which sparticle masses can be obtained 
at the LHC, an ensemble of 10,000 `gedanken experiments' were performed, 
in which for each experiment three of the gluino endpoints were 
combined with the squark endpoints obtained in \cite{Gjelsten:2004} 
in a least-square fit to give the masses. 
%
Mass differences are better determined than the masses themselves, 
a generic feature of the method. 
Furthermore, one set of endpoint measurements in general corresponds 
to several sets of masses. 
The inclusion of the gluino endpoints affects the number of minima 
and the masses of $\NO$, $\lR$, $\NT$ and $\qL$ only minimally. 
In the case of $\bO$ there is a noticeable correlation to the gluino. 
The ensemble distribution of the gluino mass was found to have a 
root-mean-square value of 7~GeV. The spread for $\mgl-\mNO$ was 
found to be about 1.5~GeV smaller. 

Spurred by the lack of significant improvement in the ensemble spread 
of $\mbO$ compared to the precision obtained for $\mgl$, 
even though the new measurements involved both, 
a sensitivity analysis for all the relevant endpoints was performed. 
This investigation showed that the three gluino endpoints being used 
have limited sensitivity to the sbottom mass, thus confirming the 
different impact they have on the gluino and sbottom masses.

Two alternative error scenarios were investigated, one in which the 
statistical errors obtained from the fast simulation of SPS~1a were doubled, 
another in which the energy scale errors were doubled. 
It was shown that the statistical errors of the endpoints, 
although smaller in magnitude than the energy scale errors, 
dominate the error on the masses. 
Only in a few exceptions, like $\mqL-\mNO$, do the 
energy scale errors play a significant role.

Finally the impact of a joint LHC--LC analysis was estimated. 
The endpoint measurements from the LHC were combined with a 
Linear Collider measurement for the LSP mass. 
This essentially fixes the SUSY mass scale. 
Consequently, in the combined analysis the masses themselves are 
determined with roughly the same precision as that of mass 
{\it differences} determined at the LHC alone. 
According to a dedicated study \cite{Allanach:2004ud}, the anticipated
accuracy will suffice to determine the high-scale mass parameters
with a precision ranging from 0.1\% for $\mHalf$ to 14\% for $\AZero$.

\appendix

\section{Calculation of $\maxmqqll$} \label{app:mqqll}
Below we use the calculation of the known squark endpoint $\maxmqFll$ 
as an aid to obtain $\maxmqqll$. 
We first follow what may seem to be the most obvious route, 
then note that while this works fine for $\maxmqFll$, the increased 
complexity of having one more particle to consider strongly limits 
the applicability to $\maxmqqll$. 
An alternative method is then developed by which $\maxmqqll$ can 
be found in a straightforward manner. 

\subsection{Method 1: By means of angles}
In the first method we keep the calculation close to the physical progression
of the decay, picking up all decay angles and eventually maximising with
respect to these.  Fig.~\ref{fig:calcmqll} shows in a stepwise manner the
$\qL$ decay chain from which $\maxmqFll$ is calculated.  Start in the rest
frame of $\qL$, and align the coordinate system to its decay products. Next,
boost to the rest frame of $\NT$, which decays such that $\lN$ is emitted at
an angle $\alpha$ relative to the direction of $\qF$. Finally, boost to the
rest frame of $\lR$, which in turn decays; for a given angle $\alpha$,
$\mqFll$ is maximised by choosing $\vvec{p}_{\lF}$ opposite to
$\vvec{p}_{\qF}+\vvec{p}_{\lN}$ in the rest frame of $\lR$ [giving a planar
decay configuration].

The resulting $\mqFll$ is expressed in terms of the angle
$\alpha$ and the four sparticle masses involved. 
In order to maximise with respect to $\alpha$, the critical
points of this constrained $\mqFll$ distribution 
must be sought 
(and tested if they are maxima or minima). 
In the part of mass
space where no critical point is found, the endpoints of the domain,
$\alpha\in[0, \pi]$, must be analysed.  In this case, three different
expressions are found for three distinct regions in mass space.  In
total, mass space is thus divided into four regions, each with its
specific expression for $\maxmqFll$, see \REFqFll.
(See \cite{Lester} for a different derivation.)

\FIGURE[ht]{
\epsfxsize 9cm
\epsfbox{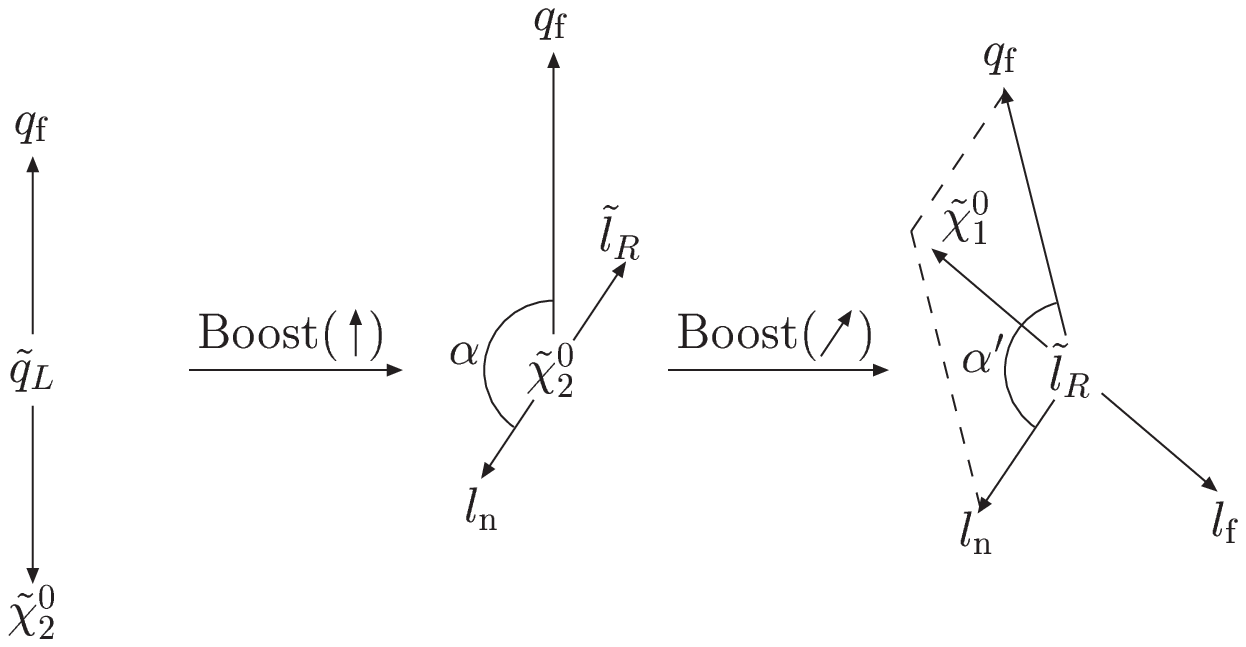}
\caption{Traditional way of calculating $\maxmqFll$.
\label{fig:calcmqll}}}

The above calculation becomes fairly involved, and is preferably 
performed with the assistance of a computer program. 
If we now want to calculate $\maxmqqll$, a gluino is added at the head of 
the chain. 
In general two more angles are needed to describe the gluino decay. 
However, since the maximum of the endpoint values are obtained with planar 
decay configurations, one of the angles can be dropped. 
The remaining additional angle is nevertheless sufficient to complicate 
significantly the search for extrema of $\mqqll$.

\subsection{Method 2: On a line}

In the alternative approach to $\maxmqFll$ (and $\maxmqqll$) 
the physical progression of the decay is not tracked. 
Instead an expression for the endpoint is found in terms of the 
one missing four-vector, that of $\NO$. 
Arguments which invoke the notion of dominant decays are then 
applied to point out a number of limiting decay configurations 
(of great geometrical simplicity) which must be considered. 

Investigating first $\maxmqFll$, 
the four-vector of the sum of the end products; the two leptons, 
the quark and the LSP, equals the four-vector of the squark. 
We therefore have
\begin{equation}
\mqL^2 = (p_{\qF ll} + p_{\NO})^2 
= \mqFll^2 + \mNO^2 + 2p_{\qF ll}\cdot p_{\NO}
\end{equation}
where $p_{\qF ll}$ is the sum of the four-vectors of the two leptons and 
the (far) quark. 
In the rest frame of $\qL$ we have
\begin{equation}
\vvec{p}_{\qF ll}=-\vvec{p}_{\NO},\quad E_{\qF ll}=\mqL-E_{\NO}
\end{equation}
which gives
\begin{eqnarray}
\mqFll^2 &=& \mqL^2 - \mNO^2 - 2\big(\mqL\sqrt{\mNO^2+\vvec{p}_{\NO}^{2}} 
- \mNO^2\big)
\end{eqnarray}
The smaller the momentum of the LSP is in the rest frame of the 
initial squark, the larger $\mqFll$ is, which agrees with one's intuition.
At the extreme, if the LSP is brought to rest in the squark rest frame, 
$\mqFll$ will attain its largest value, $\mqL-\mNO$, 
which corresponds to the critical-point solution of method 1. 

In some regions of mass space it is however not possible to have
$\NO$ at rest in the rest frame of $\qL$. 
In these regions, one of the intermediate 
sparticles is sent off with so high momentum 
that not even optimal choices of directions for the other two decays can bring 
$\NO$ to rest (in the rest frame of $\qL$). 
For the decay which gives the maximum value of $\mqFll$, 
the LSP will have a non-zero momentum in the same direction as the 
sparticle sent out from the `hardest' decay. 
There are three separate dominance regions in mass space where this happens, 
one for each sparticle decay in the cascade. 

\FIGURE[ht]{
\epsfxsize 12cm
\epsfbox{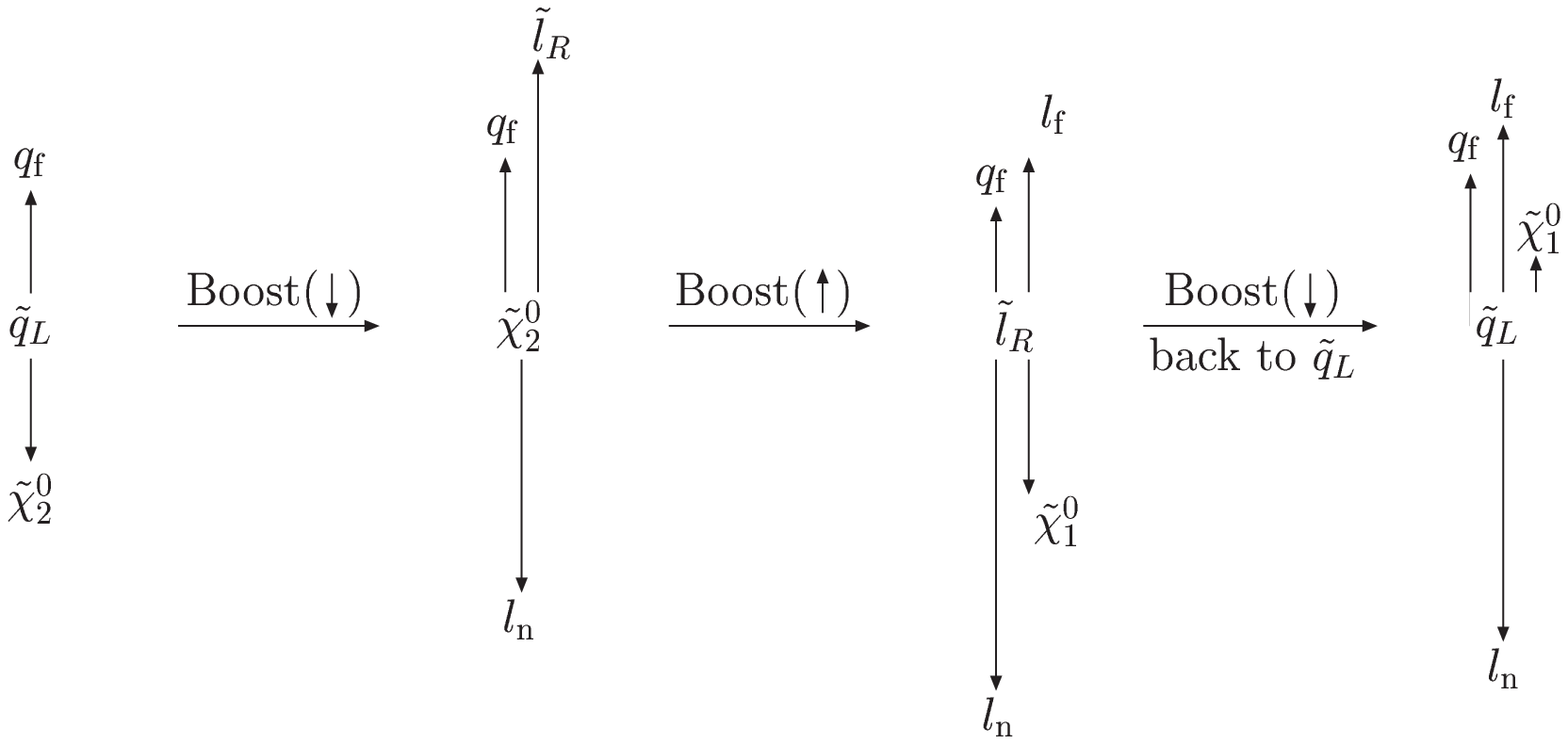}
\caption{One of three dominance regions for $\maxmqFll$
\label{fig:calcmqllcase2}}}

Consider the case where the second decay, that of $\NT$, is dominant.  If
$\NO$ cannot be brought to rest in the rest frame of $\qL$, the maximum value
is obtained for the optimised case where $\qF$ and $\lF$ fly off in one
direction and $\lN$ in the opposite, see Fig.~\ref{fig:calcmqllcase2}.  In the
last step of the figure all momenta are boosted back to the rest frame of
$\qL$. If we are inside the dominance region, $\NO$ ends up with a momentum
pointing `upwards' in the rest frame of $\qL$, as in the figure.  In this case
the defining criterion of the region takes the form,
\begin{eqnarray}
|\vvec{p}_{\lN}| > |\vvec{p}_{q} + \vvec{p}_{\lF}|
\quad\Rightarrow\quad
\rNTlR > \rlRNO\rqLNT
\end{eqnarray} 
where the three-momenta are taken in the rest frame of $\qL$. 
Note the form of the constraint in terms of {\it ratios of the masses of 
nearest neighbours in the decay chain}. The dominant ratio 
is the one which corresponds to the decay of $\NT$. 
The expression for $\maxmqFll$ in this region is then found by simply 
combining the four-momenta of the three Standard Model particles 
for the given {\em linear} configuration. 

The defining inequality of the other two dominance regions 
and the resulting expressions for $\maxmqFll$ are found 
by similar approaches. 
The general result is again \REFqFll.
Note the systematic pattern of inequalities which 
define the dominance regions. Although in cases {\itB (1)} and {\itB (3)} 
the right-hand side of the inequalities can be simplified, 
this was not done in order  
to emphasise the structure of the constraints. 

The approach described above has some clear advantages over the more 
direct approach of method 1. 
Each region is `understood' in terms of
a dominant decay, here in the sense that the mass ratio is larger than the
product of the other two, together with minimum momentum of $\NO$ in the rest
frame of $\qL$.  No angle $\alpha$ is needed. Only cascade decays on a line
need be considered, which allows the entire calculation to be performed on a
few sheets of paper.  

Furthermore, and this is the most celebrated quality here, 
the approach is trivially extendable to an arbitrary number of particles. 
In particular, the solution for $\maxmqqll$ is found in the same way, 
and is given in Eq.~(\ref{eq:edge-qqll}). 
Notice the similarity with $\maxmqFll$. 
For a detailed calculation in the fully general case of a decay chain 
involving $n$ sparticles, see~\cite{Gjelsten-thesis}.

\section{Calculation of $\maxmqqlLow$} \label{app:mqqlLow}

The calculation of $\maxmqqlLow$ turns out to be quite involved.  This is a
consequence of the inherent complexity of calculating `low'-endpoints, joined
with the many possible decay configurations available for the gluino decay
chain.  In order to demonstrate the peculiarities of `low'-endpoints in a
purer form, we first review the calculation of $\maxmqFlLow$, the simplest of
the `low'-endpoints.  For the calculation of $\maxmqqlLow$ which follow
thereafter many of the considerations are similar, only significantly more
difficult to carry through.

\subsection{Preparations: review ${\maxmqFlLow}$}\label{app:mqFlLow}

Let us first recall that since there are two leptons, two invariant
masses $\mqFl$ can be constructed, one of which will be higher than the other.
These will be denoted the `high' and `low' distributions, each of which
will have a maximum, denoted `max'.
Thus, the difficulty lies in identifying whether `high' and `low'
correspond to the `near' ($\lN$) and `far' ($\lF$) leptons, or vice
versa.  The $\mqFlHigh$ and $\mqFlLow$ distributions are constructed
from the highest/lowest of $\mqFlN$ and $\mqFlF$ on an event by event
basis.  
(We here assume that the correct jet has been selected.)
Since the $\mqFl$ value which gives the absolute maximum must
necessarily be the higher of $\mqFlN$ and $\mqFlF$ for the given
configuration, we simply have \mbox{$\maxmqFlHigh =
\max(\maxmqFlN,\maxmqFlF)$.}
For $\maxmqFlLow$ the situation is more complicated; we need to look
for the maximum value of the {\it lower} of $\mqFlN$ and $\mqFlF$.
This conditional maximisation requires that both $\mqFl$ values are
compared for the given configuration. Under no circumstance can the
endpoint be higher than the lower of the two maxima,
$\maxmqFlLow\leq\min(\maxmqFlN,\maxmqFlF)$.

Let us first assume a mass scenario where $\maxmqFlN<\maxmqFlF$, which
directly corresponds to the condition 
\begin{equation}
\mlR^2 > \mNO \mNT.
\end{equation}
This is
manifest in regions {\it (1)} and {\it (2)} of
\REFqlLowHigh.
Now, let us further assume that we
have a decay configuration where $\mqFlN$ takes on its maximum value,
$\mqFlN=\maxmqFlN$, so that $\qF$ and $\lN$ are back-to-back in the
rest frame of $\NT$.  If it is possible to choose an orientation of
$\vvec{p}_{\lF}$ such that $\mqFlF>\mqFlN$, then $\mqFlN$ can be a `low'-value,
and we will have $\maxmqFlLow=\maxmqFlN$. Specifically, this occurs
when $\mqFlF$ takes on its maximum value, with $\vvec{p}_{\lF}$ in the opposite
direction to $\vvec{p}_{\qF}$ (and parallel to $\vvec{p}_{\lN}$), 
which corresponds to 
\begin{equation}
2\slR>\sNO+\sNT,
\end{equation} i.e.\ region {\it (1)} of
\REFqlLowHigh.

If this mass constraint is not satisfied, $\mqFlN$ will be the
`high'-value for any configuration which has $\mqFlN=\maxmqFlN$ and
$\mqFlF$ will be the `low' value.
To find the configuration which gives $\maxmqFlLow$ in this case, 
consider the more general situation where the angle $\alpha$ between 
$\vvec{p}_{\qF}$ and $\vvec{p}_{\lN}$ (in the rest frame of $\NT$)
is reduced. 
The maximum value of $\mqFlF$ for this case is denoted $\maxmqFlF(\alpha)$. 
[It will occur when $\vvec{p}_{\qF}$ and $\vvec{p}_{\lF}$ are back to back 
in the $\lR$ rest frame  since $|\vvec{p}_{\lF}|$ obviously is independent 
of the orientation in this particular frame.]
If $\alpha$
is reduced (starting from $\pi$), $\mqFlN(\alpha)$ will decrease and 
$\maxmqFlF(\alpha)$ 
will increase.  As $\alpha$
decreases, the `high'-value decreases and the `low'-value increases,
and for some angle $\alpha_\equal$ they become equal:
$\mqFlN(\alpha_\equal)=\maxmqFlF(\alpha_\equal)$. If $\alpha$ is
reduced below $\alpha_\equal$, $\mqNlN$ will become the `low'-value,
and as it decreases with $\alpha$ the `low'-value will decrease.
Hence, $\alpha_\equal$ gives the optimal configuration, $\maxmqFlLow =
\mqFlN(\alpha_\equal)\equiv\maxmqFlEq$.  A simple calculation gives
\REFqlEq, 
and is valid for 
\begin{equation}
\sNO+\sNT>2\slR,
\end{equation}
i.e.\ regions {\it (2)} and {\it (3)} of 
\REFqlLowHigh.

One might expect to also find a separate solution for $\mqFlF<\mqFlN$
and $\mqFlF=\maxmqFlF$, in analogy with the first solution obtained
above. The situation is however not symmetric with respect to the two
leptons. Maximisation of $\mqFlF$ necessarily fixes $\mqFlN$ at zero,
so there is no third possibility, and once again $\maxmqFlLow =
\maxmqFlEq$. %
The general solution for $\maxmqFlLow$ is then given by two
expressions, $\maxmqFlN$ and $\maxmqFlEq$, in the mass regions derived
above, see 
\REFqlLowHigh.
(Below, we will see that a third possibility is relevant
for $\maxmqqlLow$.)

\subsection{${\maxmqqlLow}$: introduction} 
For a three-particle endpoint, the {\it configurations} which give the
maximum values $\maxmqqlN$ and $\maxmqqlF$, as well as the endpoint
expressions themselves, will be mass dependent [contrary to the
situation for $\maxmqFlN$ and $\maxmqFlF$].  In particular, we see
from 
\REFqFll\ and 
Eqs.~(\ref{eq:edge-qqlN}) and (\ref{eq:edge-qqlF}) 
of this paper that there are four different cases (mass
regions) for each.  This increases the number of situations to
consider, as well as the complexity for each.

To calculate $\maxmqqlLow$ we use the same strategy as above. 
We will first investigate the situation $\mqqlN<\mqqlF$ with 
$\mqqlN=\maxmqqlN$ (Sect.~\ref{subsect:b-first}) 
and find the appropriate conditions on the 
masses corresponding to $\maxmqqlLow=\maxmqqlN$. 
Then we will investigate the `opposite' situation, 
$\mqqlF<\mqqlN$ with $\mqqlF=\maxmqqlF$ (Sect.~\ref{subsect:b-second}), 
and find the mass conditions for the solution $\maxmqqlLow=\maxmqqlF$.
This differs from $\maxmqFlLow$, where no such solution was available. 
Finally an `equal'-solution will be sought (Sect.~\ref{subsect:b-third}). 
It will consist of a 
critical-point solution and boundary solutions (at the boundary 
of a two-dimensional parameter space). The calculation of the 
`equal'-solution is cumbersome and preferably performed with the 
aid of a computer program. 
The resulting expressions for the full solution are not very complicated, 
but many. Care should be taken in the implementation. 

In the rest frame of $\NT$ the quark sector and the lepton sector 
can be described without reference to each other. 
This makes the $\NT$ rest frame particularly convenient to use. 
Unless stated otherwise, momenta and energies are given in this 
rest frame. 
The coordinate system will be chosen such that the combined 
momentum of the two quarks, $\vvec{p}_{qq}$ will point upwards 
(along +z). 
The x and y components 
are not required, allowing the four-vectors to be given solely 
by the energy and the momentum in the z-direction.

\subsection{${\maxmqqlLow=\maxmqqlN}$\label{subsect:b-first}}
We here investigate the situation $\mqqlN<\mqqlF$ with $\mqqlN=\maxmqqlN$. 
The possible configurations with a maximised $\mqqlN$ are given in 
Table~\ref{table:mqqlN-config}.
The four rows correspond to the four different regions of mass-space
appropriate to Eq.~(\ref{eq:edge-qqlN}) (i.e.\
\REFqFll\ 
with the substitution of
Eq.~(\ref{eq:edge-qqlN})), as labeled by the first column.  The second
column shows the corresponding `region condition'. The third and fourth
column show the directions of the quark and lepton momenta, from left to
right as in the decay chain.  The combined momentum of the quarks is
always upwards (in the rest frame of $\NT$).  The direction
of $\vvec{p}_{\lN}$ is always downwards, while the direction of
$\vvec{p}_{\lF}$ is upwards for configuration~1, downwards for
configuration~2. Notice that in region {\it(4)}, configurations which
correspond to a maximum of $\mqqlLow$ do {\it not} have the quarks
aligned in the $\NT$ rest frame.
There are four mass regions and two directions for $\vvec{p}_{\lF}$, giving 
in total eight possibilities. Each must be considered. 
\begin{TABLE} {
\begin{tabular}{|c|c|c|c|}
\hline &&& \\[-4mm]
${\maxmqqlN}$
& region condition & \parbox{2.5cm}{\centering config.~1 \\ $\qN\qF\ ~\lN\lF$}& 
\parbox{2.5cm}{\centering config.~2 \\ $\qN\qF\ ~\lN\lF$}\\[3mm]
\hline &&& \\[-4mm]
region {\it (1)} & $\rglqL>\rqLNT\rNTlR$ & 
\parbox{2.5cm}{\centering $\uparrow\ \downarrow~ \downarrow\ \uparrow$ \\ superfluous} & 
\parbox{2.5cm}{\centering $\uparrow\ \downarrow~ \downarrow\ \downarrow$ \\ trivial} \\[3mm]
\hline &&& \\[-4mm]
region {\it (2)} & $\rqLNT>\rNTlR\rglqL$ &
\parbox{2.5cm}{\centering $\downarrow\ \uparrow~ \downarrow\ \uparrow$ \\ superfluous} & 
\parbox{2.5cm}{\centering $\downarrow\ \uparrow~ \downarrow\ \downarrow$ \\ trivial}\\[3mm]
\hline &&& \\[-4mm]
region {\it (3)} & $\rNTlR>\rglqL\rqLNT$ & 
\parbox{2.5cm}{\centering $\uparrow\ \uparrow~ \downarrow\ \uparrow$ \\ no solution} & 
\parbox{2.5cm}{\centering $\uparrow\ \uparrow~ \downarrow\ \downarrow$ \\ trivial}\\[3mm]
\hline &&& \\[-4mm]
region {\it (4)} & otherwise & 
\parbox{2.5cm}{\centering 
$\begin{picture}(8,10)\put(4,-2){\vector(-1,4){2.5}}\put(6,-2){\vector(1,4){2.5}}\end{picture}
\quad \downarrow\ \uparrow$ \\ superfluous} & 
\parbox{2.5cm}{\centering 
$\begin{picture}(8,10)\put(4,-2){\vector(-1,4){2.5}}\put(6,-2){\vector(1,4){2.5}}\end{picture}
\quad \downarrow\ \downarrow$ \\ trivial}\\[3mm]
\hline
\end{tabular}
\caption{Possible configurations for $\maxmqqlN$. See the text for
details.
\label{table:mqqlN-config}}
}
\end{TABLE}

The comments `trivial', `no solution' and `superfluous' give the
conclusion of the investigations. Some of these are seen right away:
in region {\it (3)}, configuration~1 the value of $\mqqlF$ evidently
vanishes since the three particles go in the same direction, hence
$\mqqlF$ must always be smaller than $\mqqlN$ and there is `no
solution'.  For configuration~2, in all four regions the condition
$\mqqlN<\mqqlF$ is simply $|\vvec{p}_{\lN}|<|\vvec{p}_{\lF}|$, which
`trivially' corresponds to
\begin{equation} \label{eq:trivial}
\sNO+\sNT<2\slR.
\end{equation}

The three regions of configuration 1 which are marked `superfluous' 
give mass conditions which turn out to be subsets of the corresponding 
(`trivial') mass condition of configuration~2
[when the appropriate region condition is also imposed]. 
This means that
if the (`superfluous') configuration~1 is possible, then also 
configuration~2 is possible.
The conclusion of this subsection will therefore be that 
$\maxmqllLow=\maxmqqlN$ if and only if the trivial condition 
(\ref{eq:trivial}) is satisfied. 
The rest of the subsection is dedicated to proving that 
the superfluous conditions are indeed contained 
in the trivial one.

For each of the three regions we must first find the mass condition 
by requiring \mbox{$\mqqlN<\mqqlF$}. 
The proofs that these conditions are already 
contained in the trivial condition 
[given that the appropriate region conditions are also imposed], 
are carried through by combining these with the `anti-trivial' condition, 
$\sNO+\sNT>2\slR$, and see that this leads to contradictions. 
For regions {\it(1)} and {\it(2)} the proofs are fairly straightforward,
but less so for region {\it(4)} due to the complicated nature of the
region condition and the non-aligned quarks. 
Below we first find the mass conditions of the three regions, then 
show the proof for one of them. 
The proofs of the two others are similar.

\subsubsection*{Regions {\it(1)} and {\it(2)}}
In the rest frame of $\NT$ the relevant four-vectors of configuration~1 are:
\begin{equation}
p_{qq}=(E_{qq},\ |\vvec{p}_{qq}|),\quad
p_{\lN}=(|\vvec{p}_{\lN}|,\ -|\vvec{p}_{\lN}|),\quad
p_{\lF}=(|\vvec{p}_{\lF}|,\ |\vvec{p}_{\lF}|)
\label{eq:pqq_N_config1}
\end{equation}
where each entry is given in the form $(E,\ p_z)$. 
The condition we need to satisfy is then:
\begin{eqnarray}
\nonumber
\mqqlN<\mqqlF &\Leftrightarrow& (p_{qq}+p_{\lN})^2<(p_{qq}+p_{\lF})^2 
\Leftrightarrow p_{qq}\cdot p_{\lN}<p_{qq}\cdot p_{\lF} 
\\
&\Leftrightarrow& 
(E_{qq}+|\vvec{p}_{qq}|)|\vvec{p}_{\lN}|<(E_{qq}-|\vvec{p}_{qq}|)|\vvec{p}_{\lF}|
\label{eq:mqqlN-massineq}
\end{eqnarray}
The lepton three-momenta have the magnitude:
\begin{eqnarray}
|\vvec{p}_{\lN}| = \frac{(\sNT-\slR)}{2\mNT},\quad
|\vvec{p}_{\lF}| = \frac{(\slR-\sNO)\mNT}{2\slR}
\label{eq:leptonmom}
\end{eqnarray}
For region {\it(1)} the quark sector gives
\begin{eqnarray}
E_{qq}=|\vvec{p}_{\qN}|+|\vvec{p}_{\qF}|,\quad && 
|\vvec{p}_{qq}|=|\vvec{p}_{\qN}|-|\vvec{p}_{\qF}|
\\
E_{qq}+|\vvec{p}_{qq}| = 2|\vvec{p}_{\qN}| = \frac{(\sgl-\sqL)\mNT}{\sqL}, \quad&& 
E_{qq}-|\vvec{p}_{qq}| = 2|\vvec{p}_{\qF}| = \frac{(\sqL-\sNT)}{\mNT}\quad
\label{eq:mqqlLow_mqqlN_E+p}
\end{eqnarray}
For region {\it(2)}, the only difference is the expression for
$|\vvec{p}_{qq}|$, which differs by an overall sign 
($|\vvec{p}_{qq}|=|\vvec{p}_{\qF}|-|\vvec{p}_{\qN}|$), and amounts to
interchanging the two expressions in (\ref{eq:mqqlLow_mqqlN_E+p}).  If
the expressions for regions {\it(1)} and {\it(2)} are inserted into
(\ref{eq:mqqlN-massineq}), the following mass conditions are found:
\begin{eqnarray}
\label{eq:mqqlLow=mqqlNcase1} \\[-1.6em]
\begin{array}[b]{lrcl}
\mbox{region {\it(1)}:}\hspace*{3mm}  & 
\slR(\sgl-\sqL)(\sNT-\slR) < \sqL(\sqL-\sNT)(\slR-\sNO) \\
\mbox{region {\it(2)}:} & 
\sqL\slR(\sqL-\sNT)(\sNT-\slR) < \mNT^4(\sgl-\sqL)(\slR-\sNO) 
\label{eq:mqqlLow=mqqlNcase2}
\end{array}
\end{eqnarray}
If (\ref{eq:mqqlLow=mqqlNcase1}) or (\ref{eq:mqqlLow=mqqlNcase2}) is satisfied, 
together with the appropriate region condition 
(Table~\ref{table:mqqlN-config}, second column), then we have 
$\maxmqqlLow=\maxmqqlN$.

\subsubsection*{Region {\it(4)}}
In region {\it(4)} the quarks are not sent out on a line
(parallel/antiparallel).  We must instead find $p_{qq}$ in the rest
frame of $\NT$ using the following equations:
\begin{eqnarray}
& (\mgl-\mlR)^2=
p_{qq\lN}^2=p_{qq}^2+2 p_{qq}\cdot p_{\lN}=
E_{qq}^2-\vvec{p}_{qq}^2+2(E_{qq}+|\vvec{p}_{qq}|)|\vvec{p}_{\lN}| &\quad
\label{eq:B9region4}
\\
& \sgl=p_{qq\NT}^2= p_{qq}^2+\mNT^2+2p_{\NT} \cdot p_{qq} =
E_{qq}^2-\vvec{p}_{qq}^2+\sNT+2\mNT E_{qq} &
\label{eq:B10region4} 
\end{eqnarray}
The first equation uses the expression for $\maxmqqlN$ found in
Eq.~(\ref{eq:edge-qqlN}). From these we find $p_{qq}$:
\begin{eqnarray}
E_{qq}=\frac{\mgl(\sNT+\slR)-2\sNT\mlR}{2\mNT\mlR},\quad
|\vvec{p}_{qq}|=\frac{\mgl(\sNT-\slR)}{2\mNT\mlR}
\\
E_{qq}+|\vvec{p}_{qq}| = \frac{\mNT(\mgl-\mlR)}{\mlR},\quad
E_{qq}-|\vvec{p}_{qq}| = \frac{(\mgl\mlR-\sNT)}{\mNT}
\end{eqnarray}
Insertion of these expressions into (\ref{eq:mqqlN-massineq}) gives 
the condition for region {\it(4)}:
\begin{eqnarray}
\mlR(\mgl-\mlR)(\sNT-\slR) < (\mgl\mlR-\sNT)(\slR-\sNO)
\label{eq:mqqlLow=mqqlNcase4}
\end{eqnarray}
If this is satisfied together with the corresponding region condition 
in the second column of Table~\ref{table:mqqlN-config}, 
then $\maxmqqlLow=\maxmqqlN$. 

However, as stated earlier the mass conditions 
(\ref{eq:mqqlLow=mqqlNcase1}), (\ref{eq:mqqlLow=mqqlNcase2}) and 
(\ref{eq:mqqlLow=mqqlNcase4}) 
are `superfluous'. 
Below this is proved for region {\itB(1)}. 
From the region condition and the `anti-trivial' condition, 
we have the following inequalities:
\begin{eqnarray}
\label{eq:B14}
\rglqL>\rqLNT\rNTlR &\Leftrightarrow& \slR(\sgl-\sqL)>\sqL(\sqL-\slR) \\
\label{eq:B15}
\sNO+\sNT>2\slR &\Leftrightarrow& (\sNT-\slR)>(\slR-\sNO)
\end{eqnarray}
Starting from (\ref{eq:mqqlLow=mqqlNcase1}) and using (\ref{eq:B14})--(\ref{eq:B15}) 
we can then write
\begin{eqnarray}
\nonumber
\sqL(\sqL-\sNT)(\slR-\sNO)
&>& \slR(\sgl-\sqL)(\sNT-\slR) \\\nonumber
&>& \sqL(\sqL-\slR)(\sNT-\slR) \\\nonumber
&>& \sqL(\sqL-\slR)(\slR-\sNO) \\
&\Rightarrow& \mlR>\mNT
\end{eqnarray}
which is clearly wrong. 
Hence, given the correct region condition, the mass condition 
(\ref{eq:mqqlLow=mqqlNcase1}) is in contradiction to 
the anti-trivial condition, meaning that it must be 
contained in the trivial condition. 
Similar proofs apply for the two other superfluous cases. 

We therefore have the following conclusion, as already stated: 
if (\ref{eq:trivial}) is satisfied, then $\maxmqllLow=\maxmqqlN$.

\subsection{${\maxmqqlLow=\maxmqqlF}$\label{subsect:b-second}}
%
\noindent
Here we investigate the situation \mbox{$\mqqlF<\mqqlN$} with
$\mqqlF=\maxmqqlF$.  The possible configurations with a maximised
$\mqqlF$ are given in Table~\ref{table:mqqlF-config} (in contrast to
Table~\ref{table:mqqlN-config} where $\mqqlN$ was maximised). 
As before we consider the rest frame of $\NT$ and align the coordinate 
system such as to have $\vvec{p}_{qq}$ pointing upwards. 
The maximum value for $\mqqlF$ is found for the decay configuration of $\NT$ 
and $\lR$ which maximises $\vvec{p}_{\lF}$ downwards 
(in the rest frame of $\NT$). 
This is achieved if first $\lR$, then $\lF$, are emitted downwards. 
Since this necessarily fixes $\vvec{p}_{\lN}$ upwards, there is only one 
configuration in Table~\ref{table:mqqlF-config}.
The mass ratios in the quark sector are
the same for $\maxmqqlF$ as for $\maxmqqlN$, but in the lepton sector
the relevant ratio is now \mbox{$\mNT/(\mNO\mNT/\mlR) = \mlR/\mNO$,}
see (\ref{eq:edge-qqlN}) and (\ref{eq:edge-qqlF}), which changes
the region conditions in column 2 as compared to those of
Table~\ref{table:mqqlN-config}.

\begin{TABLE} { \hspace{2cm}
\begin{tabular}{|c|c|c|}
\hline && \\[-4mm]
${\maxmqqlF}$
& region condition & \parbox{2.5cm}{\centering config.~1 \\ $\qN\qF\ ~ \lN\lF$} \\[3mm]
\hline && \\[-4mm]
region {\it (1)} & $\rglqL>\rqLNT\rlRNO$ &
\parbox{2.5cm}{\centering $\uparrow\ \downarrow~ \uparrow\ \downarrow$ \\ unique} \\[3mm]
\hline && \\[-4mm]
region {\it (2)} & $\rqLNT>\rlRNO\rglqL$ &
\parbox{2.5cm}{\centering $\downarrow\ \uparrow~ \uparrow\ \downarrow$ \\ unique} \\[3mm]
\hline && \\[-4mm]
region {\it (3)} & $\rlRNO>\rglqL\rqLNT$ &
\parbox{2.5cm}{\centering $\uparrow\ \uparrow~ \uparrow\ \downarrow$ \\ no solution} \\[3mm]
\hline && \\[-4mm]
region {\it (4)} & otherwise &
\parbox{2.5cm}{\centering 
$\begin{picture}(8,10)\put(4,-2)
{\vector(-1,4){2.5}}\put(6,-2){\vector(1,4){2.5}}\end{picture}
\quad \uparrow\ \downarrow $ \\ unique} \\[3mm]
\hline 
\end{tabular} \hspace{2cm}
\caption{Possible configurations for $\maxmqqlF$. See the text for details. 
\label{table:mqqlF-config}}
}
\end{TABLE}

Region{\it (3)} has no solution since $\mqqlN$ vanishes and therefore
cannot be the `high'-value.  In the three other regions there are unique
mass conditions for the solution $\maxmqqlLow=\maxmqqlF$, which is
what we seek here.  
The calculations follow the exact same path as in the previous subsection. 

\subsubsection*{Regions {\it(1)} and {\it(2)}}
In the rest frame of $\NT$ the relevant four-vectors are 
[compare with (\ref{eq:pqq_N_config1})]:
\begin{equation}
p_{qq}=(E_{qq},\ |\vvec{p}_{qq}|),\quad
p_{\lN}=(|\vvec{p}_{\lN}|,\ |\vvec{p}_{\lN}|),\quad
p_{\lF}=(|\vvec{p}_{\lF}|,\ -|\vvec{p}_{\lF}|)
\end{equation}
The condition we need to satisfy is now:
\begin{eqnarray}
\nonumber
\mqqlF<\mqqlN &\Leftrightarrow& (p_{qq}+p_{\lF})^2<(p_{qq}+p_{\lN})^2 
\Leftrightarrow p_{qq}\cdot p_{\lF}<p_{qq}\cdot p_{\lN} 
\\
&\Leftrightarrow& 
(E_{qq}+|\vvec{p}_{qq}|)|\vvec{p}_{\lF}|<(E_{qq}-|\vvec{p}_{qq}|)|\vvec{p}_{\lN}|
\label{eq:mqqlF-massineq}
\end{eqnarray}
The lepton momenta are still given by (\ref{eq:leptonmom}). 
The energy and momentum of the quarks are the same as in 
the previous subsection:
Eq.~(\ref{eq:mqqlLow_mqqlN_E+p}) for region~{\itB(1)}, 
the expressions interchanged for region~{\itB(2)}.

When these expressions 
are inserted into 
(\ref{eq:mqqlF-massineq}), we find the following mass conditions: 
\begin{eqnarray}
\label{eq:mqqlLow=mqqlFcase1} \\[-1.6em]
\begin{array}[b]{lrcl}
\mbox{region {\it(1)}:}\hspace*{3mm}  & \mNT^4(\sgl-\sqL)(\slR-\sNO) 
&<& \sqL\slR(\sqL-\sNT)(\sNT-\slR)\quad \\
\mbox{region {\it(2)}:} & \sqL(\sqL-\sNT)(\slR-\sNO) &<& \slR(\sgl-\sqL)(\sNT-\slR)
\label{eq:mqqlLow=mqqlFcase2}
\end{array}
\end{eqnarray}
If (\ref{eq:mqqlLow=mqqlFcase1}) or (\ref{eq:mqqlLow=mqqlFcase2}) 
is satisfied, 
together with the appropriate region condition 
(Table~\ref{table:mqqlF-config}, second column), then we have 
$\maxmqqlLow=\maxmqqlF$.

\subsubsection*{Region {\it(4)}}
In region {\it(4)} 
the combination of 
\begin{eqnarray}
& (\mgl-\mNO\mNT/\mlR)^2=
p_{qq\lF}^2=p_{qq}^2+2 p_{qq}\cdot p_{\lF}=
E_{qq}^2-\vvec{p}_{qq}^2+2(E_{qq}+|\vvec{p}_{qq}|)|\vvec{p}_{\lF}| &\qquad
\end{eqnarray}
[equivalent to (\ref{eq:B9region4})] with (\ref{eq:B10region4})
results in: 
\begin{eqnarray}
E_{qq}=\frac{\mgl(\slR+\sNO)-2\mNT\mlR\mNO}{2\mlR\mNO},\quad
|\vvec{p}_{qq}|=\frac{\mgl(\slR-\sNO)}{2\mlR\mNO}
\\
E_{qq}+|\vvec{p}_{qq}| = \frac{(\mgl\mlR-\mNT\mNO)}{\mNO},\quad
E_{qq}-|\vvec{p}_{qq}| = \frac{(\mgl\mNO-\mNT\mlR)}{\mlR}
\end{eqnarray}
Insertion of these expressions into (\ref{eq:mqqlF-massineq}) gives 
the condition for region {\it(4)}:
\begin{eqnarray}
\sNT(\mgl\mlR-\mNT\mNO)(\slR-\sNO)
< \mlR\mNO(\mgl\mNO-\mNT\mlR)(\sNT-\slR)\qquad 
\label{eq:mqqlLow=mqqlFcase4}
\end{eqnarray}
If this is satisfied together with the corresponding region condition, 
then $\maxmqqlLow=\maxmqqlF$. 

This result completes the special-case solutions of $\maxmqqlLow$.

\subsection{${\maxmqqlLow=\maxmqqlEq}$\label{subsect:b-third}}

If none of the above mass conditions are fulfilled, the maximum 
of $\mqqlLow$ will in general be reached in an `equal-solution' 
decay configuration, where $\mqqlN=\mqqlF$. 

The relevant information on the two first steps of the cascade decay
is contained in the value of $\pqq$.  (We stay in the rest frame of
$\NT$.)  Its maximum value is attained if $\qN$ and $\qF$ are sent off
in the same direction, and is given by,
\begin{equation}
\pqqP=\frac{\sgl-\sNT}{2\mNT} 
\label{eq:app-pqqP}
\end{equation}
If sent off in opposite directions, $\pqq$ is at its minimum,
\begin{equation}
\pqqAP=\frac{|\sgl\sNT-\mqL^4|}{2\sqL\mNT}
\label{eq:app-pqqAP}
\end{equation}
In the following we let $\pqq$ be a free variable and only later
impose the physical constraint $\pqq\in[\pqqAP,\pqqP]$.

As before, we align the coordinate system to have $\pqqS$ along +z.
The first lepton, $\lN$, is emitted at an angle $\alpha$ relative to
$\pqqS$.  The direction of the second lepton, $\lF$, is chosen to maximise 
$\mqqlF$ for the given $\alpha$, allowing the notation $\maxmqqlF(\alpha)$.  
We have this freedom since we are interested 
in the maximum, not just any $\mqqlEq$ value.  At this stage the decay
is parametrised in terms of two variables, $\pqq$ and $\alpha$.  By
requiring $\mqqlN(\alpha)=\maxmqqlF(\alpha)$, the angle 
can be expressed in terms of $\pqq$:
\begin{equation}
\label{eq:mqqlLow-equal-cosbeta}
\cos\alpha_{\equal} = 
\frac{\slR\big(\sNO+\sNT-2\slR\big)\big(\sqrt{\sgl+\pqq^2}-\mNT\big)
-\mNT\big(\slR-\sNO\big)\varphi(\pqq)}
{\big(\sNT-\slR\big)\big(2\slR-\sNO\big)\pqq}
\end{equation}
where $\varphi(\pqq)$ is given by
\begin{eqnarray}
\varphi(\pqq) &=& 
\left[\sNT\pqq^2 +\left(\sNO+\sNT-2\slR\right)\left(\sgl+\sNT-2\mNT\sqrt{\sgl+\pqq^2}\right)
\right]^{1/2} \nonumber \\
\end{eqnarray}
[Later we will need to ensure that $|\cos\alpha_{\equal}|\leq1$.]
Insertion of this solution into $\mqqlN(\alpha)$ [or $\maxmqqlF(\alpha)$]
returns an expression for $\mqqlEq$ in terms of $\pqq$ only.  Since we
are interested in a maximum value we now search for the critical point
of $\mqqlEq$, and find an expression for the critical momentum,
\begin{eqnarray}
\pqqCrit&=&
\left[
\left(
2\mlR(\sNO+\sNT-2\slR)+(3\slR-\sNO)\sqrt{\sgl+2\slR-\sNO-\sNT}\ 
\right)^2 \right. \nonumber
\\&&
\qquad - 4\sgl\sNT\slR 
\Big]^\half
\frac{1}{2\mNT\mlR}
\label{eq:app-pqqCrit}
\end{eqnarray}
Inserting this back into $\mqqlEq$ gives an expression for the
critical point solution,

\begin{eqnarray}
\nonumber
\left(\mqqlEqCrit\right)^2 &=&
\sgl - \left[ \mNT(3\slR-\sNO)\sqrt{\pqqCrit^2+\sgl} \right. \\
&& \qquad \left.
-\sNT\slR-(\slR-\sNO)\varphi(\pqqCrit)
\right]
\big/ (2\slR-\sNO)
\label{eq:app-mqqlEqCrit}
\end{eqnarray}
All expressions found from the procedure above are formal.  In order
for them to be physical, the momentum and the angle $\alpha$ of the
critical solution must lie in the allowed regions:
\begin{eqnarray}
\label{eq:EqCritCondpqq}
\pqqCrit &\in& [\pqqAP,\pqqP]
\\
\label{eq:EqCritCondCosB}
\cos\alpha_{\equal}(\pqqCrit) &\in& [-1,1]
\end{eqnarray}
For a given set of masses 
Eqs.~(\ref{eq:EqCritCondpqq}) and (\ref{eq:EqCritCondCosB}) 
must be tested numerically for the resulting $\pqqCrit$. 

If the critical solution is {\it not} physical, the maximum must lie
on the boundary of the two-dimensional domain defined by
Eqs.~(\ref{eq:EqCritCondpqq}) and (\ref{eq:EqCritCondCosB}).  
For $\cos\alpha_{\equal}=\pm1$ the appropriate $|\vvec{p}_{qq}|$ can be 
found by redoing the procedure leading to 
(\ref{eq:app-mqqlEqCrit}) but with $\alpha$ fixed at 0 or $\pi$. 
This gives the following expression for $|\vvec{p}_{qq}|$: 
\begin{eqnarray}
\nonumber
\pqqCosB&=&
\frac{\pm(\sNO\sNT-\mlR^4)}{4\mNT\slR(\sNT-\slR)(\slR-\sNO)}
\Big[-(2\sNT\slR-\sNO\sNT-\mlR^4)
\label{eq:app-pqqCosBeq1}
\\
&&
+\sqrt{(\sNO\sNT-\mlR^4)^2 + 4\sgl\slR(\sNT-\slR)(\slR-\sNO)}
\Big]
\end{eqnarray}
The invariant mass can then be found straightforward by taking 
$m_{qq\lN}^2=(p_{qq}+p_{\lN})^2$ 
and using (\ref{eq:B10region4}). 
This gives
\begin{eqnarray}
\big(\mqqlEqCosB\big)^2 &=&
\Big[(\sgl+\slR)\mNT - (\sNT+\slR)\sqrt{\pqqCosB^2+\sgl} 
\nonumber\\&&
\mp 
(\sNT-\slR)\pqqCosB\Big]
/\mNT
\label{eq:app-mqqlEqCosBeq1}
\end{eqnarray}

The boundary solutions at maximum and minimum $|\vvec{p}_{qq}|$ are most 
easily found by inserting the boundary values 
(\ref{eq:app-pqqP})--(\ref{eq:app-pqqAP}) into the general solution 
(\ref{eq:app-mqqlEqCrit}). 
This gives 
\begin{eqnarray}
\big(\mqqlEqP\big)^2 &=&
(\sgl-\sNT)(\slR-\sNO)/(2\slR-\sNO)
\label{eq:app-mqqlEqP}
\\
\nonumber
\big(\mqqlEqAP\big)^2 &=&
\big[-(3\slR-\sNO)(\sgl\sNT+\mqL^4) + 2\sqL[\sgl(2\slR-\sNO)+\sNT\slR]
\\
&&
+2\sqL(\slR-\sNO)\varphi(\pqqAP)\big]
/[2\sqL(2\slR-\sNO)]
\label{eq:app-mqqlEqAP}
\end{eqnarray}

All expressions (\ref{eq:app-mqqlEqCosBeq1})--(\ref{eq:app-mqqlEqAP}) 
are formal. For a given set of masses one must explicitly (numerically)
impose the one of Eqs.~(\ref{eq:EqCritCondpqq}) and
(\ref{eq:EqCritCondCosB}) which is not satisfied by construction.

\subsection{General solution}

Putting all this together, the fully general solution for $\maxmqqlLow$ is
given by
\begin{eqnarray}
\maxmqqlLow = 
\left\{  
\begin{array}{ll}
\maxmqqlN   &\ {\rm for}\quad 
\sNO+\sNT<2\slR \\[2mm]
\maxmqqlF   &\ {\rm for}\quad \av{\rm cut 1} \\[2mm]
\maxmqqlEq  &\ {\rm otherwise}\quad
\end{array}
\right.
\label{eq:edge-qqlLow}
\end{eqnarray}
where
\begin{eqnarray}
\nonumber
\displaystyle
\begin{array}{lclcll}
\av{\rm cut 1} \; = && \displaystyle
\Bigg[
\rglqL>\rqLNT\rlRNO 
&\bigwedge& \displaystyle
\frac{\mNT^4 \left(\sgl-\sqL\right)\left(\slR-\sNO\right)}
{\sqL\slR\left(\sqL-\sNT\right)\left(\sNT-\slR\right)} <1  & \Bigg] \\[6mm]
&\bigvee& \displaystyle
\Bigg[ \rqLNT>\rlRNO\rglqL 
&\bigwedge& \displaystyle
\frac{\sqL\left(\sqL-\sNT\right)\left(\slR-\sNO\right)}
{\slR\left(\sgl-\sqL\right)\left(\sNT-\slR\right)} <1  & \Bigg] \\[6mm]
&\bigvee& \displaystyle
\Bigg[ \rglqL<\rqLNT\rlRNO 
&\bigwedge& 
\multicolumn{2}{l}{{\displaystyle \rqLNT<\rlRNO\rglqL}
\bigwedge
{\displaystyle \rlRNO <\rglqL\rqLNT}} \\[6mm]
&&
&\bigwedge& 
\multicolumn{2}{l}{\displaystyle
\frac{\sNT\left(\mgl\mlR-\mNT\mNO\right)\left(\slR-\sNO\right)}
{\mlR\mNO\left(\mgl\mNO-\mNT\mlR\right)\left(\sNT-\slR\right)}<1
 \Bigg]}
\end{array} \\
\label{eq:cut1}
\end{eqnarray}
is constructed from (\ref{eq:mqqlLow=mqqlFcase1}), 
(\ref{eq:mqqlLow=mqqlFcase2}) and (\ref{eq:mqqlLow=mqqlFcase4}) 
together with the appropriate region conditions. 
Expressions for $\maxmqqlN$ and $\maxmqqlF$ are given in
Eqs.~(\ref{eq:edge-qqlN}) and (\ref{eq:edge-qqlF}) respectively,
and $\maxmqqlEq$ is given by
\begin{eqnarray}
\maxmqqlEq &=& 
\max\left(\mqqlEqCrit,\mqqlEqCosB,\mqqlEqP,\mqqlEqAP\right)
\end{eqnarray}
which uses (\ref{eq:app-mqqlEqCrit}) and 
(\ref{eq:app-mqqlEqCosBeq1})--(\ref{eq:app-mqqlEqAP}).

\acknowledgments
This work has been performed partly within the ATLAS Collaboration,
and we thank collaboration members for helpful discussions.
We have made use of the physics analysis framework and tools
which are the results of collaboration-wide efforts.
We are in particular grateful to Giacomo Polesello for numerous discussions.
BKG would like to thank Steinar Stapnes for useful discussions. 
This research has been supported in part by the Research Council of Norway.


\end{document}